\documentclass[
aps,%
prl,%
reprint, 
twocolumn,%
groupedaddress,%
superscriptaddress,%
amsfonts,%
footinbib,%
floatfix,%
longbibliography
]{revtex4-2}

\usepackage[utf8]{inputenc} 
\usepackage[T1]{fontenc} 
\usepackage{lmodern} %
\usepackage[english]{babel}

\usepackage[pdftex]{hyperref}
\hypersetup{
    colorlinks=true,
    citecolor= magenta,
    linkcolor= blue,
    filecolor=blue,      
    urlcolor= blue,
    breaklinks= true,
    pdfauthor = {Saranyo Moitra},
    pdfcreator = {\LaTeX\ and \flqq hyperref\frqq},
}

\usepackage{orcidlink}

\usepackage[pdftex,dvipsnames]{xcolor} 
\usepackage{graphicx} 

\definecolor{maroon}{RGB}{126, 0, 0}
\definecolor{olive}{RGB}{0, 126, 0}
\definecolor{teal}{RGB}{74,113,139}

\usepackage{booktabs} 
\usepackage{array} 

\usepackage{enumitem} 

\usepackage{amsmath,amsfonts,amssymb,amsthm}
\usepackage{mathtools} 
\usepackage{nicematrix}
\usepackage[light,bbsymbols]{bboldx} 
\DeclareSymbolFontAlphabet{\mathbb}{AMSb}
\DeclareSymbolFontAlphabet{\mathbbl}{bboldx}

\usepackage{comment}
\usepackage{xargs} 

\usepackage[colorinlistoftodos,prependcaption,textsize=tiny,textwidth=1.4cm]{todonotes}
\newcommandx{\highlightnote}[5][1=]{\textcolor{#3}{#4}\todo[linecolor=#3!25,backgroundcolor=#3!25,bordercolor=#3,#1]{{\fontfamily{lmss}\textbf{#2}:\newline \selectfont #5}}}

\newcommandx{\sm}[3][1=, 3=check]{
\highlightnote[#1]{Saranyo}{red}{#2}{#3}
}
\newcommandx{\inti}[3][1=, 3=check]{
\highlightnote[#1]{Inti}{blue}{#2}{#3}
}
\newcommandx{\editor}[4][1=, 2=editor, 4=check]{
\highlightnote[#1]{#2}{Plum}{#3}{#4}
}

\usepackage[color,final]{showkeys}
\definecolor{refkey}{gray}{.75}
\definecolor{labelkey}{RGB}{192,0,0} 

\newcommand{\ii}{\iota}
\newcommand{\e}{\,\mathrm{e}}

\newcommand{\ve}[1]{{\boldsymbol{#1}}}
\newcommand{\cmplx}[1]{\mathsf{#1}}

\newcommand{\tx}[1]{{\text{#1}}}

\newcommand{\qk}{\ve{\mathrm{q}}}

\newcommand{\rr}{\ve{\mathrm{r}}}
\newcommand{\rp}{{\ve{\mathrm{r}}'}}
\newcommand{\xx}{\ve{\mathrm{x}}}

\newcommand{\zee}{\cmplx{z}}

\newcommand{\eps}{\epsilon}

\newcommand{\flux}{\Phi}

\newcommand{\lb}{\ell_B}

\newcommand{\avg}[1]{#1_{\mathrm{ca}}}

\usepackage{physics} 

\usepackage[capitalise]{cleveref}

\graphicspath{ {./Figs/} }

\begin{document}

\title{
Instability of Laughlin FQH liquids into gapless power-law correlated states with continuous exponents in ideal Chern bands: rigorous results from plasma mapping
}

\author{Saranyo Moitra\,\orcidlink{0000-0001-7912-1961}}
\email{saranyo.moitra@uni-leipzig.de}

\author{Inti \surname{Sodemann Villadiego}\,\orcidlink{0000-0002-1824-5167}}
\email{sodemann@uni-leipzig.de}
\affiliation{
Institut f\"ur Theoretische Physik,
Universit\"at Leipzig,
04103 Leipzig, Germany
}

\begin{abstract}
We investigate the fate of Laughlin's wave-function in ideal Chern bands which can be mapped to generalized zero Landau levels in spatially dependent magnetic fields. By exploiting its exact mapping onto a classical Coulomb gas and leveraging previous results of one-component plasmas in nonuniform neutralizing backgrounds, we demonstrate that the ideal Laughlin wave-function undergoes a phase transition from its well-known fully gapped topologically ordered plasma state into a power-law correlated dielectric state even for the fixed filling of $1/3$, as the magnetic field becomes increasingly more inhomogeneous. This dielectric state is gapless even though it does not spontaneously break translational symmetry. Remarkably, for a fixed filling $\nu=1/m$, the exponent governing density correlations in this state changes continuously as a function of the degree of spatial inhomogeneity of the magnetic field, and can range from $4$ near a Berezinskii-Kosterlitz-Thouless transition to the plasma state, up to $2 m$ in the limit of fields generated by point solenoids. 
\end{abstract}

\date{\today}

\maketitle

\textcolor{blue}{\emph{Introduction.}}
A curious property of 2D Dirac fermions \cite{Aharonov.Casher_PRA79_GroundStateSpintextonehalf} and parabolic fermions with $g$-factor $g=2$ \cite{Dubrovin.Novikov_JETP80_GroundStatesTwodimensional}, is that their zero-energy states in a magnetic field remain exactly degenerate at zero energy even when the field is spatially non-uniform. These wavefunctions span a generalized zero Landau level that can be viewed as an ideal Chern band \cite{Roy_PRB14_BandGeometryFractional,Jackson.Roy_NC15_GeometricStabilityTopological,Claassen.Devereaux_PRL15_PositionMomentumDualityFractional,Tarnopolsky.Vishwanath_PRL19_OriginMagicAngles,Ledwith.Vishwanath_PRR20_FractionalChernInsulator,Wang.Yang_PRL21_ExactLandauLevel,Ozawa.Mera_PRB21_RelationsTopologyQuantum,Mera.Ozawa_PRB21_KahlerGeometryChern,Ledwith.Parker_PRB23_VortexabilityUnifyingCriterion,Estienne.Crepel_PRR23_IdealChernBands}. Such ideal Chern bands (ICBs) have been found in the chiral limit of twisted bilayer graphene \cite{Tarnopolsky.Vishwanath_PRL19_OriginMagicAngles,Ledwith.Vishwanath_PRR20_FractionalChernInsulator}, argued to approximate the bands of twisted MoTe$_2$ \cite{Morales-Duran.MacDonald_PRL24_MagicAnglesFractional,Shi.MacDonald_PRB24_AdiabaticApproximationAharonovCasher,Li.Wu_PRB25_VariationalMappingChern} and to be relevant to rhombohedral multilayer graphene \cite{Tan.Devakul_PRX24_ParentBerryCurvature,Dong.Parker_PRL24_AnomalousHallCrystals,Tan.Devakul_PRL25_VariationalWaveFunctionAnalysis}, where fractional quantum Hall effect at zero magnetic field has been  observed \cite{Cai.Xu_N23_SignaturesFractionalQuantum,Zeng.Shan_N23_ThermodynamicEvidenceFractional,Park.Xu_N23_ObservationFractionallyQuantized,Xu.Li_PRX23_ObservationIntegerFractional,Lu.Ju_N24_FractionalQuantumAnomalous}.

These ICBs, also referred to as Aharonov-Casher (AC) bands \cite{Morales-Duran.MacDonald_PRL24_MagicAnglesFractional}, admit an explicit writing of a generalized Laughlin wave-function \cite{Ledwith.Vishwanath_PRR20_FractionalChernInsulator}. However, unlike the standard zero Landau level where the Laughlin wavefunctions are fixed, the ICBs realize a continuum {\it family} of Laughlin wavefunctions, which can be viewed as parametrized by a spatially dependent magnetic field $B(\rr)$ \footnote{In general this field is a function that parametrizes single-particle wavefunctions and not necessarily a physical field.}. The central question of our study is what is the nature of the physical states realized by these generalized Laughlin wave-functions? as we will see, in ICBs the Laughlin wave-function can realize not only the well-known topologically ordered state with fractionalized quasiparticles, but also a variety of other complex states. We will demonstrate this by exploiting the exact mapping of the probability density of the Laughlin wavefunction onto a classical Coulomb gas \cite{Laughlin_PRL83_AnomalousQuantumHall}, which has been recently used to investigate optical properties of these states \cite{Wolf.Su_PRL25_IntrabandCollectiveExcitations}. In this plasma mapping, the electrons act as point-like $-1$ charges fluctuating at temperature $T=1/2m$ (at filling $\nu=1/m$), and the magnetic field as a fixed neutralizing background with charge density $\rho_\mathrm{b}(\rr)=B(\rr)/(m\flux_0)$.

In this setting, the standard topologically ordered phase of the Laughlin wave-function is a classical plasma where correlations decay exponentially due to its perfect screening properties. However, as we will see, in ICBs the spatial variations of the neutralizing background tend to localize the electrons,  transforming the Laughlin plasma into a dielectric \cite{Clerouin.Piller_PRA87_TwodimensionalClassicalElectron,Alastuey.Jancovici_PRA88_CommentTwodimensionalClassical,Choquard.Hansen_PRA89_IonizationPhaseDiagram}. In the case of the ICBs relevant to moire systems, the unit cell contains one effective flux quantum, and, therefore, the neutralizing background has fractional charge $1/m$ per unit cell. In these situations, the dielectric localization of electrons is expected to be accompanied by spontaneous breaking of discrete translations, i.e. charge-density-wave (CDW) formation with $m$-fold enlargement of the unit cell. 

However, as we will demonstrate, these dielectrics realized by the Laughlin wave-function are not ordinary CDWs, but rather complex states with density correlations that decay as power laws with continuously changing exponents. Such decay of correlations implies that these states are gapless, but this gaplessness does not seem to fit within the Goldstone mode paradigm because the continuous translational symmetry has been broken explicitly by the spatially dependent field. To be able to tread more carefully through these complexities, we will focus on a simpler model that deviates from the situation in moire materials and has $m$-fluxes per unit cell, so that each cell contains $+1$ background charge and the dielectric tends to localize one electron per unit cell. A state with one electron strongly localized per lattice site, e.g. by a deep single-particle potential, is expected to be adiabatically connected to a trivial insulator described by a Slater determinant. However, as we will see, the dielectric Laughlin state with one electron per unit cell generically remains in a complex gapless and power-law correlated state with continuously varying exponents.

\textcolor{blue}{\emph{Model and preliminaries: }} 
A simple ICB is realized by massless two-dimensional Dirac fermions placed in an inhomogeneous magnetic field $\ve{B}=-B(\rr)\,\ve{\rm \hat{z}}$, which have the following zero-energy   wavefunctions \cite{Aharonov.Casher_PRA79_GroundStateSpintextonehalf}:
    \begin{equation}\label{eq:ACstate}
        \psi_n(\rr)=\zee^n\e^{-\phi(\rr)},\quad n=0,1,2,\dots,
    \end{equation}
here we use the Coulomb gauge $\nabla \cdot A(\rr)=0$, $\zee=x+\ii y$ is the complex position, and the exponent is determined from $\laplacian{\phi(\rr)}=eB(\rr)/\hbar$ (electron charge is $-e$, $e>0$). The above wave-functions are normalizable when  $B(\rr)$ has a positive global spatial average, but we will focus on cases when $B(\rr)$ is spatially periodic.
%
The generalized Laughlin state of $N$ electrons is given by:
    \begin{equation}\label{eq:Gen_laughlin}
        \Psi(\rr_1,\ldots,\rr_N)=\prod_{i<j}(\zee_i-\zee_j)^{m}\prod_{i=1}^{N}\e^{-\phi(\rr_i)}.
    \end{equation}
For a uniform field, $B(\rr)=B_0$, $\phi(\rr)=|\rr|^2/4\lb^2, \lb=\sqrt{\hbar/eB_0}$,  Eq.~\eqref{eq:Gen_laughlin} reduces to the usual Laughlin state. As in the usual zero Landau level, the above wave-function corresponds to the highest-degree holomorphic polynomial that is a zero-energy eigen-state of a short-ranged Hamiltonian \cite{Haldane_PRL83_FractionalQuantizationHall,Trugman.Kivelson_PRB85_ExactResultsFractional}, whose projection onto the ICB we view as the ideal Hamiltonian defining our problem. Remarkably, the probability density of these generalized Laughlin states is identical to the finite temperature probability density of a gas of classical particles repelling via 2D Coulomb forces and with a neutralizing background charge density, namely $|\Psi|^2 = \exp[-\beta \mathcal{U}]$, with:
 \begin{equation}\label{eq:U_genlaughlin}
        \mathcal{U}=
        -\sum_{i<j}^{N} \ln |\rr_i-\rr_j|
        +
        \sum_{i=1}^{N} \frac{\phi(\rr_i)}{m},
    \end{equation}
\noindent where we have chosen the electrons to have Coulomb charge $-1$ and $\beta=2m$. The term $\phi(\rr_i)/m$ is the potential energy experienced by an electron at $\rr_i$ generated by a background with charge density $\rho_\mathrm{b}(\rr)=B(\rr)/(m\flux_0)$, where $\flux_0=h/e$. Thus, we see that any region, $R$, containing a single flux quantum, $\int_R d^2 r B(\rr)=\Phi_0$, acts as a background source of charge $+1/m$. 

In usual zero Landau levels, the above is the map to a one-component Coulomb gas with a uniform neutralizing background $\rho_\mathrm{b}(\rr)=1/(2m\pi\lb^2)\equiv B_0/(m\flux_0)$, introduced by Laughlin \cite{Laughlin_PRL83_AnomalousQuantumHall}. This one-component gas is in a plasma phase at high temperatures (the quantum topologically ordered state) and undergoes a phase transition into a crystal at a very large exponent $m\gtrsim 70$ \cite{Caillol.Hansen_JSP82_MonteCarloStudy,Herland.Sudbo_PRB13_FreezingUnconventionalTwodimensional}. However, from numerical \cite{Clerouin.Piller_PRA87_TwodimensionalClassicalElectron,Choquard.Hansen_PRA89_IonizationPhaseDiagram} and analytical studies \cite{Alastuey.Jancovici_PRA88_CommentTwodimensionalClassical} of the one-component Coulomb gas in a non-uniform  neutralizing backgrounds, we expect that even the most common Laughlin state with $m=3$  undergoes a transition from its plasma state into a dielectric state where the electrons largely freeze their positions, when the spatial variations of the background charge density (i.e. of $B(\rr)$) are sufficiently large. 

When the magnetic field magnetic field $B(\rr)$ contains one flux quantum per unit cell, as realized in Chern bands of some moiré materials (see e.g. \cite{Wu.MacDonald_PRL19_TopologicalInsulatorsTwisted} for twisted MoTe$_2$), there will be an average of $1/m$ electrons per unit cell, and the Laughlin dielectric state would tend to break spontaneously the discrete magnetic translational symmetries defined by $B(\rr)$. However, we will see that such dielectric state is not a trivial CDW state (i.e. one approximated by a Slater determinant), but rather a complex power-law correlated state that remains gapless even in the absence of spontaneous symmetry breaking. 

\begin{figure}[t]
    \centering
    \includegraphics[width=\columnwidth]{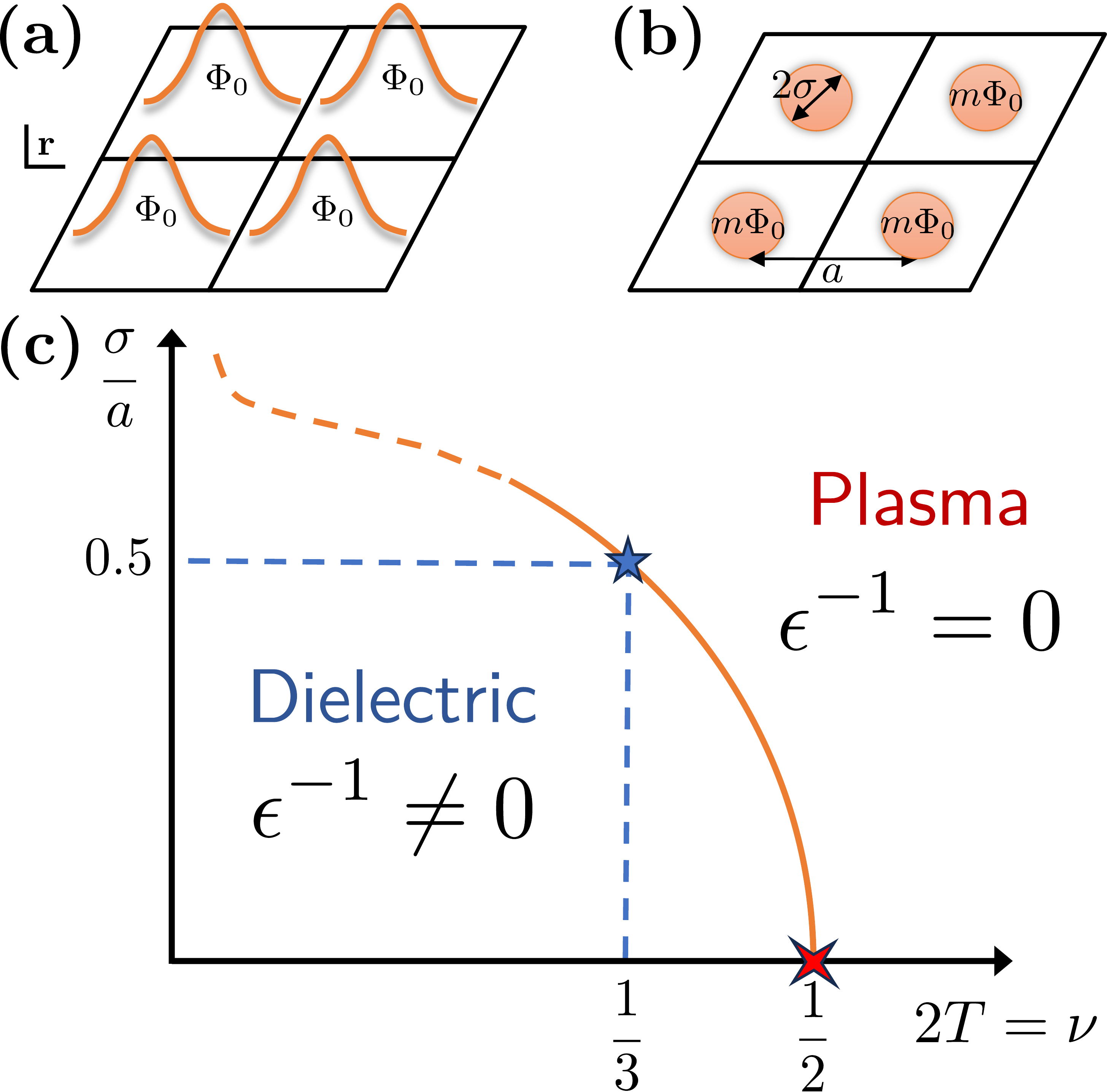}
    \caption{
    \textbf{(a)} Depiction of effective magnetic field in moiré Chern bands with one flux per unit cell, and  
    \textbf{(b)} in ICB of solenoids with $m$-fluxes per unit cell. 
    \textbf{(c)} Phase diagram of the Coulomb gas with neutralizing background of the ions from panel \textbf{(b)}, adapted from Ref.~\cite{Choquard.Hansen_PRA89_IonizationPhaseDiagram}. The BKT transition is expected to occur for small radius and at $T\eps=1/4$.}
    \label{fig:phasediagram}
\end{figure}

To argue more clearly for this, we adopt a setting where it is manifest that such dielectric Laughlin state does not need to spontaneously break  translational symmetry. To do so, we take $B(\rr)$ to be a periodic function with $m$-fluxes per unit cell. In this case, the neutrality of the Coulomb gas implies that we must have an average of one electron per unit cell, and therefore the dielectric can form without spontaneously breaking translations. For concreteness, we take $B(\rr)$ to be uniform inside a collection of solenoids of radius $\sigma$ and zero outside the solenoids. The solenoid positions, $\ve{\rm X}_i$, are taken to form a triangular lattice (i.e. with separation $a= (4\pi m/\sqrt{3})^{1/2}l_B$, see Fig.\ref{fig:phasediagram}(b)). None of these details are crucial for our main conclusions, but this geometry has the advantage that it has been previously investigated numerically~\cite{Choquard.Hansen_PRA89_IonizationPhaseDiagram} and analytically \cite{Alastuey.Jancovici_PRA88_CommentTwodimensionalClassical}. Since each solenoid carries flux $m\Phi_0$, it behaves as a fixed disk with positive charge $+1$, and, therefore, this Coulomb gas is a 2D version of a classical crystal of ``Thomson plum pudding'' of hydrogen atoms. The Coulomb potential in Eq.\eqref{eq:U_genlaughlin} originating from these positive background ions is a sum of the single ion potentials (denoted by $\chi(\rr)$):
 \begin{equation}\label{eq:solenoidflux}
        \frac{\phi(\rr)}{m}=
        \sum_{i}\chi(\rr-\ve{\rm X}_i), \ 
        \chi(\rr)=\begin{dcases}
            \frac{|\rr|^2-\sigma^2}{2\sigma^2} & r<\sigma
            \\
            \ln\frac{|\rr|}{\sigma} & r>\sigma
        \end{dcases}
        .
    \end{equation}

This model has two dimensionless parameters: the dimensionless solenoid radius, $\sigma/a$, and the filling (or temperature) $\nu=1/m=2 T$, and its phase diagram, adapted from  Ref.~\cite{Choquard.Hansen_PRA89_IonizationPhaseDiagram}, is depicted in Fig.\ref{fig:phasediagram}(c). Several arguments \cite{Alastuey.Jancovici_PRA88_CommentTwodimensionalClassical} support that in the ``dilute'' limit of point-like solenoids ($\sigma/a \ll 1$), the high-temperature plasma state with ionized electrons (i.e. the quantum topologically ordered phase) undergoes a Berezinskii–Kosterlitz–Thouless (BKT) transition into a dielectric state in which each background positive ion binds one electron, at the temperature that corresponds to the filling of the bosonic Laughlin state at $\nu=1/2$ (red star in Fig.\ref{fig:phasediagram}(c)). For the fermionic Laughlin state at $\nu=1/3$, from the results of Ref.~\cite{Choquard.Hansen_PRA89_IonizationPhaseDiagram}, we estimate that the plasma-dielectric transition occurs at $\sigma/a \sim 0.5$ (blue star in Fig.\ref{fig:phasediagram}(c)), i.e. when the solenoids are barely touching.


\textcolor{blue}{\emph{Dielectric properties and quasi-holes: }} The dielectric constant $\epsilon$ plays a central role on the properties of the Coulomb gas. In the plasma state $\epsilon^{-1}=0$, while in the dielectric $\epsilon$ is finite and changes continuously as a function of the solenoid radius, $\sigma/a$, and temperature $T=1/(2m)$. At the BKT transition $\epsilon^{-1}$ experiences a universal jump from $\epsilon^{-1}=4T=2/m$ to $\epsilon^{-1}=0$. More generally, the screening of external potentials is controlled by the dielectric response function, $\eps^{-1}(\rr,\rp)$, which is  related to the density correlator $S(\rr,\rr')$ \cite{Alastuey.Jancovici_PRA88_CommentTwodimensionalClassical,Martin.Gruber_JSP83_NewProofStillingerLovett} (see \cite{suppl} for details), as follows:
\begin{equation}\label{eq:eps_rrp}
        \eps^{-1}(\rr,\rp)=\delta(\rr-\rp)
        +2m\int\dd{\rr''}\ln\frac{|\rr-\rr''|}{\sigma}\,S(\rr'',\rp),
    \end{equation}
\noindent where $S(\rr,\rp)\coloneq\ev{\rho(\rr)\rho(\rp)}-\ev{\rho(\rr)}\ev{\rho(\rp)}$, and $\rho(\rr)=\sum_{i=1}^{N}\delta(\rr-\rr_i)$ is the electron density. In our setting, these correlation functions are only invariant under the discrete lattice translations of the periodic magnetic field, and thus, to facilitate the analysis of their long-distance behaviors, it is convenient to define their \emph{cell-averaged} counterparts as follows:
\begin{equation}
    \begin{aligned}
        \avg{\eps^{-1}}(\rr)
        &\coloneq
            \frac{1}{A_\tx{uc}}\int_{\xx\in\tx{uc}}\dd[2]{\xx} \,\eps^{-1}(\rr+\xx,\xx),
    \end{aligned}
    \end{equation}
\noindent where $A_\tx{uc}$ is the area of a unit cell and we define similarly $\avg{S}(\rr)$. The Fourier transform of the cell-averaged Eq.~\eqref{eq:eps_rrp} is
    $
        \avg{\eps^{-1}}(\qk)=1-4\pi m\,\avg{S}(\qk)/|\qk|^2
    $, and the dielectric constant $\eps$ is obtained from its limit $\qk\to0$ \cite{Alastuey.Jancovici_PRA88_CommentTwodimensionalClassical,Martin.Gruber_JSP83_NewProofStillingerLovett}:
    \begin{equation}\label{eq:eps_from_Sca}
        \eps^{-1}=1- \lim_{\qk\to0}\frac{4\pi m\avg{S}(\qk)}{|\qk|^2}=1+\pi m\int\dd[2]{\rr}|\rr|^2\avg{S}(\rr),
    \end{equation}
\noindent The above equations assume that the system has at least $C_3$ rotational invariance, and imply that  $\avg{S}(\rr)$ must decay faster than $\sim 1/|\rr|^4$ at long distances ($1/|\rr|^4$ is the decay at the BKT line). The above reduces to the Stillinger-Lovett  perfect screening condition \cite{Mitchell.Groeneveld_JSP77_SecondmomentConditionStillinger,Martin.Gruber_JSP83_NewProofStillingerLovett,Martin_RMP88_SumRulesCharged} 
in the plasma state. From it one can also relate the  dielectric constant to the variance of particle positions \cite{Herland.Sudbo_PRB13_FreezingUnconventionalTwodimensional,suppl} as follows ($\rr_\tx{CM}\equiv(\sum_i\rr_i)/N$):
\begin{equation}\label{eq:eps_rCM}
    \eps^{-1}=1-\frac{2m\pi }{A_\tx{uc}}N(\ev{\rr_\tx{CM}^2}-\ev{\rr_\tx{CM}}^2).
\end{equation} 
%

Now, the probabilities of our Coulomb gas with an additional point-like test particle located at the origin and charge $-1/m$, are identical to the probability density of the Laughlin quasi-hole wave-function \cite{Laughlin_PRL83_AnomalousQuantumHall} generalized to ICB, i.e. $\Psi_\tx{qh}=(\prod_{i=1}^N \zee_i)\,\Psi$, with $\Psi$ the wave-function from Eq.\eqref{eq:Gen_laughlin}. This quasi-hole is also an exact zero-energy eigen-state of the same pseudo-potential Hamiltonian projected into the ICB, for which $\Psi$ is a zero-energy eigen-state. At long distances, the Coulomb electric field emanating from this test charge would be reduced by the dielectric constant $\epsilon$. In the plasma phase there is perfect screening ($\epsilon^{-1}=0$), and this test charge is surrounded by an exponentially localized screening cloud which exactly cancels its bare charge, leading to the classic result that quasi-hole carries a fractional physical electric charge $q_\tx{qh}=+e/m$ (i.e. a depletion of electrons). In the dielectric state $\epsilon$ is finite, and the screening is imperfect, and the electric field far away from the test particle asymptotes to the value  $-\hat{\rr}/(\epsilon m |\rr|)$. From this it follows that the quasi-hole in the dielectric state carries a physical electric charge:
 \begin{equation}\label{eq:qhcharge}
        q_\tx{qh}=(1-\eps^{-1})\frac{e}{m}.
    \end{equation}
\noindent The above is remarkable because in the dielectric state $\epsilon$ changes continuously with the solenoid radius, even for fixed $m$, and therefore the quasi-hole charge changes continuously. This quasi-hole is a non-local object because the decay of its screened electric field with distance implies that electrons have a displacement away from the solenoid centers that decays as $\sim 1/ |\rr|$ away from the location of the quasi-hole. These features indicate that the dielectric Laughlin state cannot be a fully gapped phase of matter, since in 2D these are expected to have quasi-particles with physical charges that are fixed rational values of the electron charge \cite{Wen_07_QuantumFieldTheory}. We will provide a more rigorous argument for this based on the behavior of correlation functions in the coming section.

\textcolor{blue}{\emph{Correlations and cluster expansion.}} 
In this section we will demonstrate one of the most remarkable properties of the dielectric  Laughlin states in ICBs, namely, that their density-density correlations have the following power-law decay at long distances: 
\begin{equation}\label{eq:Sdecay}
S(\rr,\rp)\sim -\frac{C}{|\rr-\rp|^{2m \eps^{-1}}}.
\end{equation}
\noindent Notably, the exponent $2m\epsilon^{-1}$ changes continuously in the dielectric state, even for a fixed $m$, as a function of the solenoid radius $\sigma$.   
Since gapped ground states of local Hamiltonians necessarily have exponentially decaying correlations \cite{Hastings.Koma_CMP06_SpectralGapExponential}, the above implies that the dielectric Laughlin states must have gapless excitations, even if they do not spontaneously break translations. The above decay of density correlations has been  widely discussed for the dielectric state of the more commonly studied two-component Coulomb gas \cite{Berezinskii_SPJ71_DestructionLongrangeOrder,Berezinskii_SPJ72_DestructionLongrangeOrder,Kosterlitz.Thouless_JPCSSP73_OrderingMetastabilityPhase,Alastuey.Cornu_JSP92_CorrelationsKosterlitzThoulessPhase,Alastuey.Forrester_JSP95_CorrelationsTwocomponentLoggas,minnhagen1987two}, but also occurs in the one-component Coulomb gas with nonuniform background \cite{Alastuey.Jancovici_PRA88_CommentTwodimensionalClassical}. 
Here, we will reproduce these results by following the cluster expansion method proposed in Ref.\cite{Alastuey.Jancovici_PRA88_CommentTwodimensionalClassical} for our one-component Coulomb gas with nonuniform background.

Deep in the dielectric phase ($\sigma/a\ll 1, m^{-1}\ll 1$) each positive ion strongly binds one electron. This motivates a cluster expansion controlled by these small parameters~\cite{Alastuey.Jancovici_PRA88_CommentTwodimensionalClassical}. We start by an exact re-writing of the Boltzmann weight from Eq.~\eqref{eq:U_genlaughlin} as follows:
\begin{equation}\label{eq:cluster}
    \e^{-2m\mathcal{U}}=\prod_{i=1}^{N}p(\rr_i-\ve{\rm X}_i) \prod_{j<k}(1+f_{jk})
\end{equation}
where $p(\rr)=\e^{-2m\chi(\rr)}/w$, $w=\int\dd[2]{\rr}\e^{-2m\chi(\rr)}\simeq\order{\sigma^2}$ is the probability density of a single electron and a single ion (i.e. as if no other ions or electrons were present), and the Mayer factors are $f_{jk}\coloneq\exp[-2m\,\delta\mathcal{U}_{jk}]-1$, with the residual potential defined as 
$
\delta\mathcal{U}_{jk}=
-\ln(|\rr_{j}-\rr_k|/\sigma)
-\chi(\ve{\rm X}_j-\ve{\rm X}_k)
+\chi(\rr_j-\ve{\rm X}_k)
+\chi(\rr_k-\ve{\rm X}_j)
$. We have introduced a few innocuous constants in the definition of $\mathcal{U}$ to facilitate its normalization  \cite{suppl}. The cluster expansion is a perturbation series in $f_{jk}$, justified from the smallness of the residual potential, $\delta\mathcal{U}_{jk}$, when particles and ions are tightly bound. To leading order ($f_{jk}\approx 0$) one gets 
$S(\rr,\rp)\approx \sum_i p(\rr-\ve{\rm X}_i)(\delta(\rr-\rp)-p(\rp-\ve{\rm X}_i))$,
whose cell average decays at long distances, as follows \cite{suppl}:
    \begin{align}\label{eq:Sca_powerlaw}
        \avg{S}(\rr)
        &\approx -\frac{1+P_\tx{in}}{A_\tx{uc}}\,p(\rr)
        = -\frac{1+P_\tx{in}}{A_\tx{uc}w}\,\frac{\sigma^{2m}}{|\rr|^{2m}}, \ |\rr|\gg\sigma,
    \end{align}
where $P_\tx{in}=\int_{r<\sigma}\dd[2]{\rr}p(\rr)$ is the ($\order{1}$) probability that a single particle is inside a solenoid.
\noindent The above proves Eq.\eqref{eq:Sdecay} in the dilute limit ($\sigma/a \to 1$) where one expects $\epsilon^{-1}\to 1$, and is consistent with a BKT transition into the plasma state at $\beta=2m=4$ ~\cite{Alastuey.Jancovici_PRA88_CommentTwodimensionalClassical}. Within this leading order, one finds $N\ev{\rr_\tx{CM}^2}=\int\dd[2]{\rr}r^2\,p(\rr)$, and from Eq.~\eqref{eq:eps_rCM}, $\eps$ is estimated to be: 
\begin{equation}\label{eq:epsinverse}
        \eps^{-1}=1- 
        \frac{(m-1)(m-2+(m+2)e^{-m})}
        {(m-2)(m-1+e^{-m})} 
        \frac{2\pi\sigma ^2}{A_\tx{uc}}.
    \end{equation}
\noindent Notice that $\eps^{-1}$ remains finite as the temperature approaches zero ($m\to\infty$) and one recovers the polarizability of a Thomson atom from Ref.~\cite{Choquard.Hansen_PRA89_IonizationPhaseDiagram}
(Ref.~\cite{Alastuey.Jancovici_PRA88_CommentTwodimensionalClassical} considered the ions to be impenetrable). The first-order in $f_{jk}$ correction to $\avg{S}(\rr)$, are expected to lead to the following modifications at large distances \cite{Alastuey.Jancovici_PRA88_CommentTwodimensionalClassical}:
    \begin{equation}\label{SwithLog}
        \avg{S}(\rr)\sim \frac{\sigma^{2m}}{|\rr|^{2m}}
        \left[
        1+2m(1-\eps^{-1})\ln\frac{|\rr|}{\sigma}
        \right],
    \end{equation}
where $\eps^{-1}$ is given by Eq.~\eqref{eq:epsinverse}. By noting that $1-\eps^{-1}$ is of order $\sigma^2/a^2$, 
Eq.~\eqref{SwithLog} can be re-exponentiated in the small $\sigma/a$ limit to read $\avg{S}(\rr)\approxeq\left(\sigma/|\rr|\right)^{2m \eps^{-1}}$ \cite{Alastuey.Jancovici_PRA88_CommentTwodimensionalClassical}. This decay of density-density correlations also implies that the density of the screening cloud surrounding the quasi-hole decays as $\delta n(\rr)\sim A/|\rr|^{2(m\epsilon^{-1}-1)}$, and therefore this is sufficiently fast to make the quasi-hole charge from Eq.\eqref{eq:qhcharge} well defined in the dielectric state since $m\epsilon^{-1}>2$ (except at the BKT line where $m\epsilon^{-1} \rightarrow 2$).

\textcolor{blue}{\emph{Discussion.}}
We have used a Coulomb gas mapping to investigate the nature of the Laughlin wave-function in ideal Chern bands which can be viewed as generalized zero Landau levels with non-uniform magnetic fields. By taking a model of a field generated by a lattice of solenoids carrying $m$-fluxes, we have demonstrated that the Laughlin wave-function is not always in its celebrated plasma state with topological order, but can transition into a new dielectric state where electrons tend to localize and bind to these solenoids. This dielectric Laughlin state displays power law decaying correlations with continuous exponents, which implies that the quantum state must be gapless, but this gaplessness appears to be beyond the standard Goldstone mode paradigm, since there is no continuous translational symmetry in the model. 

Quantum ground states with continuously varying exponents in two space dimensions are quite rare. However, similar behavior has been  reported in a  supersymmetric XY model \cite{rana1993soluble}, and in quantum-dimer and related models \cite{Ardonne.Fradkin_AoP04_TopologicalOrderConformal,Fradkin.Sondhi_PRB04_BipartiteRokhsarKivelsonPoints,Vishwanath.Senthil_PRB04_QuantumCriticalityDeconfinement,isakov2011dynamics}, which have been argued to be critical states with continuously varying exponents captured by a 2D quantum Lifshitz field theory   \cite{Ardonne.Fradkin_AoP04_TopologicalOrderConformal,Hsu.Fradkin_PRB13_DynamicalStabilityQuantum,Vishwanath.Senthil_PRB04_QuantumCriticalityDeconfinement}. Thus, it is possible that the dielectric Laughlin states in ICBs might form a similar type of critical states. We hope future studies can further clarify this interesting possibility and investigate their response to perturbations, as well as the mechanisms underlying their gaplessness which appears to be beyond Goldstone mode paradigm.

\textcolor{blue}{\emph{Acknowledgements.}} We would like to thank Qi Hu, Markus Kohler, Anish Koley, Akihiro Ozawa, Nemin Wei, Luca Delacr\'etaz, Dam T. Son, Paul Wiegmann, Nicol\'as Morales Dur\'an, Gil Young Cho, Jennifer Cano, Lei Chen, Trithep Devakul, Cenke Xu, Daniel Arovas, Ying Ran, Ashvin Vishwanath, Debanjan Chowdhury, Ravindra N. Bhatt, Ganpathy Murthy, Daniel Barci and Jörg Schmalian for stimulating discussions and pointing out useful references. ISV is also thankful for the hospitality at the Aspen Center for Physics, supported by National Science Foundation (NSF) grant PHY-2210452, where some of these discussions took place. We are also thankful for the support from the Deutsche Forschungsgemeinschaft (DFG) through research grants Project No. 542614019, No. 518372354, No. 555335098.

\bibliography{LaughlinCrystal}

\begin{thebibliography}{2}%
\makeatletter
\providecommand \@ifxundefined [1]{%
 \@ifx{#1\undefined}
}%
\providecommand \@ifnum [1]{%
 \ifnum #1\expandafter \@firstoftwo
 \else \expandafter \@secondoftwo
 \fi
}%
\providecommand \@ifx [1]{%
 \ifx #1\expandafter \@firstoftwo
 \else \expandafter \@secondoftwo
 \fi
}%
\providecommand \natexlab [1]{#1}%
\providecommand \enquote  [1]{``#1''}%
\providecommand \bibnamefont  [1]{#1}%
\providecommand \bibfnamefont [1]{#1}%
\providecommand \citenamefont [1]{#1}%
\providecommand \href@noop [0]{\@secondoftwo}%
\providecommand \href [0]{\begingroup \@sanitize@url \@href}%
\providecommand \@href[1]{\@@startlink{#1}\@@href}%
\providecommand \@@href[1]{\endgroup#1\@@endlink}%
\providecommand \@sanitize@url [0]{\catcode `\\12\catcode `\$12\catcode
  `\&12\catcode `\#12\catcode `\^12\catcode `\_12\catcode `\%12\relax}%
\providecommand \@@startlink[1]{}%
\providecommand \@@endlink[0]{}%
\providecommand \url  [0]{\begingroup\@sanitize@url \@url }%
\providecommand \@url [1]{\endgroup\@href {#1}{\urlprefix }}%
\providecommand \urlprefix  [0]{URL }%
\providecommand \Eprint [0]{\href }%
\providecommand \doibase [0]{https://doi.org/}%
\providecommand \selectlanguage [0]{\@gobble}%
\providecommand \bibinfo  [0]{\@secondoftwo}%
\providecommand \bibfield  [0]{\@secondoftwo}%
\providecommand \translation [1]{[#1]}%
\providecommand \BibitemOpen [0]{}%
\providecommand \bibitemStop [0]{}%
\providecommand \bibitemNoStop [0]{.\EOS\space}%
\providecommand \EOS [0]{\spacefactor3000\relax}%
\providecommand \BibitemShut  [1]{\csname bibitem#1\endcsname}%
\let\auto@bib@innerbib\@empty
\bibitem [{\citenamefont {Martin}\ and\ \citenamefont
  {Gruber}(1983)}]{Martin.Gruber_JSP83_NewProofStillingerLovett}%
  \BibitemOpen
  \bibfield  {author} {\bibinfo {author} {\bibfnamefont {{\relax Ph}.~A.}\
  \bibnamefont {Martin}}\ and\ \bibinfo {author} {\bibfnamefont {{\relax
  Ch}.}~\bibnamefont {Gruber}},\ }\bibfield  {title} {\bibinfo {title} {A new
  proof of the {{Stillinger-Lovett}} complete shielding condition},\ }\href
  {https://doi.org/10.1007/BF01019506} {\bibfield  {journal} {\bibinfo
  {journal} {J Stat Phys}\ }\textbf {\bibinfo {volume} {31}},\ \bibinfo {pages}
  {691} (\bibinfo {year} {1983})}\BibitemShut {NoStop}%
\bibitem [{\citenamefont {Martin}(1988)}]{Martin_RMP88_SumRulesCharged}%
  \BibitemOpen
  \bibfield  {author} {\bibinfo {author} {\bibfnamefont {{\relax Ph}.~A.}\
  \bibnamefont {Martin}},\ }\bibfield  {title} {\bibinfo {title} {Sum rules in
  charged fluids},\ }\href {https://doi.org/10.1103/RevModPhys.60.1075}
  {\bibfield  {journal} {\bibinfo  {journal} {Rev. Mod. Phys.}\ }\textbf
  {\bibinfo {volume} {60}},\ \bibinfo {pages} {1075} (\bibinfo {year}
  {1988})}\BibitemShut {NoStop}%
\end{thebibliography}%


\begin{thebibliography}{52}%
\makeatletter
\providecommand \@ifxundefined [1]{%
 \@ifx{#1\undefined}
}%
\providecommand \@ifnum [1]{%
 \ifnum #1\expandafter \@firstoftwo
 \else \expandafter \@secondoftwo
 \fi
}%
\providecommand \@ifx [1]{%
 \ifx #1\expandafter \@firstoftwo
 \else \expandafter \@secondoftwo
 \fi
}%
\providecommand \natexlab [1]{#1}%
\providecommand \enquote  [1]{``#1''}%
\providecommand \bibnamefont  [1]{#1}%
\providecommand \bibfnamefont [1]{#1}%
\providecommand \citenamefont [1]{#1}%
\providecommand \href@noop [0]{\@secondoftwo}%
\providecommand \href [0]{\begingroup \@sanitize@url \@href}%
\providecommand \@href[1]{\@@startlink{#1}\@@href}%
\providecommand \@@href[1]{\endgroup#1\@@endlink}%
\providecommand \@sanitize@url [0]{\catcode `\\12\catcode `\$12\catcode
  `\&12\catcode `\#12\catcode `\^12\catcode `\_12\catcode `\%12\relax}%
\providecommand \@@startlink[1]{}%
\providecommand \@@endlink[0]{}%
\providecommand \url  [0]{\begingroup\@sanitize@url \@url }%
\providecommand \@url [1]{\endgroup\@href {#1}{\urlprefix }}%
\providecommand \urlprefix  [0]{URL }%
\providecommand \Eprint [0]{\href }%
\providecommand \doibase [0]{https://doi.org/}%
\providecommand \selectlanguage [0]{\@gobble}%
\providecommand \bibinfo  [0]{\@secondoftwo}%
\providecommand \bibfield  [0]{\@secondoftwo}%
\providecommand \translation [1]{[#1]}%
\providecommand \BibitemOpen [0]{}%
\providecommand \bibitemStop [0]{}%
\providecommand \bibitemNoStop [0]{.\EOS\space}%
\providecommand \EOS [0]{\spacefactor3000\relax}%
\providecommand \BibitemShut  [1]{\csname bibitem#1\endcsname}%
\let\auto@bib@innerbib\@empty
\bibitem [{\citenamefont {Aharonov}\ and\ \citenamefont
  {Casher}(1979)}]{Aharonov.Casher_PRA79_GroundStateSpintextonehalf}%
  \BibitemOpen
  \bibfield  {author} {\bibinfo {author} {\bibfnamefont {Y.}~\bibnamefont
  {Aharonov}}\ and\ \bibinfo {author} {\bibfnamefont {A.}~\bibnamefont
  {Casher}},\ }\bibfield  {title} {\bibinfo {title} {Ground state of a
  spin-{$\frac{1}{2}$} charged particle in a two-dimensional magnetic field},\
  }\href {https://doi.org/10.1103/PhysRevA.19.2461} {\bibfield  {journal}
  {\bibinfo  {journal} {Phys. Rev. A}\ }\textbf {\bibinfo {volume} {19}},\
  \bibinfo {pages} {2461} (\bibinfo {year} {1979})}\BibitemShut {NoStop}%
\bibitem [{\citenamefont {Dubrovin}\ and\ \citenamefont
  {Novikov}(1980)}]{Dubrovin.Novikov_JETP80_GroundStatesTwodimensional}%
  \BibitemOpen
  \bibfield  {author} {\bibinfo {author} {\bibfnamefont {A.}~\bibnamefont
  {Dubrovin}}\ and\ \bibinfo {author} {\bibfnamefont {S.~P.}\ \bibnamefont
  {Novikov}},\ }\bibfield  {title} {\bibinfo {title} {Ground states of a
  two-dimensional electron in a periodic magnetic field},\ }\href
  {http://jetp.ras.ru/cgi-bin/e/index/e/52/3/p511?a=list} {\bibfield  {journal}
  {\bibinfo  {journal} {Sov. {{Phys}}. {{JETP}}}\ }\textbf {\bibinfo {volume}
  {52}},\ \bibinfo {pages} {511} (\bibinfo {year} {1980})}\BibitemShut
  {NoStop}%
\bibitem [{\citenamefont {Roy}(2014)}]{Roy_PRB14_BandGeometryFractional}%
  \BibitemOpen
  \bibfield  {author} {\bibinfo {author} {\bibfnamefont {R.}~\bibnamefont
  {Roy}},\ }\bibfield  {title} {\bibinfo {title} {Band geometry of fractional
  topological insulators},\ }\href {https://doi.org/10.1103/PhysRevB.90.165139}
  {\bibfield  {journal} {\bibinfo  {journal} {Phys. Rev. B}\ }\textbf {\bibinfo
  {volume} {90}},\ \bibinfo {pages} {165139} (\bibinfo {year}
  {2014})}\BibitemShut {NoStop}%
\bibitem [{\citenamefont {Jackson}\ \emph {et~al.}(2015)\citenamefont
  {Jackson}, \citenamefont {M{\"o}ller},\ and\ \citenamefont
  {Roy}}]{Jackson.Roy_NC15_GeometricStabilityTopological}%
  \BibitemOpen
  \bibfield  {author} {\bibinfo {author} {\bibfnamefont {T.~S.}\ \bibnamefont
  {Jackson}}, \bibinfo {author} {\bibfnamefont {G.}~\bibnamefont
  {M{\"o}ller}},\ and\ \bibinfo {author} {\bibfnamefont {R.}~\bibnamefont
  {Roy}},\ }\bibfield  {title} {\bibinfo {title} {Geometric stability of
  topological lattice phases},\ }\href {https://doi.org/10.1038/ncomms9629}
  {\bibfield  {journal} {\bibinfo  {journal} {Nat Commun}\ }\textbf {\bibinfo
  {volume} {6}},\ \bibinfo {pages} {8629} (\bibinfo {year} {2015})}\BibitemShut
  {NoStop}%
\bibitem [{\citenamefont {Claassen}\ \emph {et~al.}(2015)\citenamefont
  {Claassen}, \citenamefont {Lee}, \citenamefont {Thomale}, \citenamefont
  {Qi},\ and\ \citenamefont
  {Devereaux}}]{Claassen.Devereaux_PRL15_PositionMomentumDualityFractional}%
  \BibitemOpen
  \bibfield  {author} {\bibinfo {author} {\bibfnamefont {M.}~\bibnamefont
  {Claassen}}, \bibinfo {author} {\bibfnamefont {C.~H.}\ \bibnamefont {Lee}},
  \bibinfo {author} {\bibfnamefont {R.}~\bibnamefont {Thomale}}, \bibinfo
  {author} {\bibfnamefont {X.-L.}\ \bibnamefont {Qi}},\ and\ \bibinfo {author}
  {\bibfnamefont {T.~P.}\ \bibnamefont {Devereaux}},\ }\bibfield  {title}
  {\bibinfo {title} {Position-{{Momentum Duality}} and {{Fractional Quantum
  Hall Effect}} in {{Chern Insulators}}},\ }\href
  {https://doi.org/10.1103/PhysRevLett.114.236802} {\bibfield  {journal}
  {\bibinfo  {journal} {Phys. Rev. Lett.}\ }\textbf {\bibinfo {volume} {114}},\
  \bibinfo {pages} {236802} (\bibinfo {year} {2015})}\BibitemShut {NoStop}%
\bibitem [{\citenamefont {Tarnopolsky}\ \emph {et~al.}(2019)\citenamefont
  {Tarnopolsky}, \citenamefont {Kruchkov},\ and\ \citenamefont
  {Vishwanath}}]{Tarnopolsky.Vishwanath_PRL19_OriginMagicAngles}%
  \BibitemOpen
  \bibfield  {author} {\bibinfo {author} {\bibfnamefont {G.}~\bibnamefont
  {Tarnopolsky}}, \bibinfo {author} {\bibfnamefont {A.~J.}\ \bibnamefont
  {Kruchkov}},\ and\ \bibinfo {author} {\bibfnamefont {A.}~\bibnamefont
  {Vishwanath}},\ }\bibfield  {title} {\bibinfo {title} {Origin of {{Magic
  Angles}} in {{Twisted Bilayer Graphene}}},\ }\href
  {https://doi.org/10.1103/PhysRevLett.122.106405} {\bibfield  {journal}
  {\bibinfo  {journal} {Phys. Rev. Lett.}\ }\textbf {\bibinfo {volume} {122}},\
  \bibinfo {pages} {106405} (\bibinfo {year} {2019})}\BibitemShut {NoStop}%
\bibitem [{\citenamefont {Ledwith}\ \emph {et~al.}(2020)\citenamefont
  {Ledwith}, \citenamefont {Tarnopolsky}, \citenamefont {Khalaf},\ and\
  \citenamefont
  {Vishwanath}}]{Ledwith.Vishwanath_PRR20_FractionalChernInsulator}%
  \BibitemOpen
  \bibfield  {author} {\bibinfo {author} {\bibfnamefont {P.~J.}\ \bibnamefont
  {Ledwith}}, \bibinfo {author} {\bibfnamefont {G.}~\bibnamefont
  {Tarnopolsky}}, \bibinfo {author} {\bibfnamefont {E.}~\bibnamefont
  {Khalaf}},\ and\ \bibinfo {author} {\bibfnamefont {A.}~\bibnamefont
  {Vishwanath}},\ }\bibfield  {title} {\bibinfo {title} {Fractional {{Chern}}
  insulator states in twisted bilayer graphene: {{An}} analytical approach},\
  }\href {https://doi.org/10.1103/PhysRevResearch.2.023237} {\bibfield
  {journal} {\bibinfo  {journal} {Phys. Rev. Res.}\ }\textbf {\bibinfo {volume}
  {2}},\ \bibinfo {pages} {023237} (\bibinfo {year} {2020})}\BibitemShut
  {NoStop}%
\bibitem [{\citenamefont {Wang}\ \emph {et~al.}(2021)\citenamefont {Wang},
  \citenamefont {Cano}, \citenamefont {Millis}, \citenamefont {Liu},\ and\
  \citenamefont {Yang}}]{Wang.Yang_PRL21_ExactLandauLevel}%
  \BibitemOpen
  \bibfield  {author} {\bibinfo {author} {\bibfnamefont {J.}~\bibnamefont
  {Wang}}, \bibinfo {author} {\bibfnamefont {J.}~\bibnamefont {Cano}}, \bibinfo
  {author} {\bibfnamefont {A.~J.}\ \bibnamefont {Millis}}, \bibinfo {author}
  {\bibfnamefont {Z.}~\bibnamefont {Liu}},\ and\ \bibinfo {author}
  {\bibfnamefont {B.}~\bibnamefont {Yang}},\ }\bibfield  {title} {\bibinfo
  {title} {Exact {{Landau Level Description}} of {{Geometry}} and
  {{Interaction}} in a {{Flatband}}},\ }\href
  {https://doi.org/10.1103/PhysRevLett.127.246403} {\bibfield  {journal}
  {\bibinfo  {journal} {Phys. Rev. Lett.}\ }\textbf {\bibinfo {volume} {127}},\
  \bibinfo {pages} {246403} (\bibinfo {year} {2021})}\BibitemShut {NoStop}%
\bibitem [{\citenamefont {Ozawa}\ and\ \citenamefont
  {Mera}(2021)}]{Ozawa.Mera_PRB21_RelationsTopologyQuantum}%
  \BibitemOpen
  \bibfield  {author} {\bibinfo {author} {\bibfnamefont {T.}~\bibnamefont
  {Ozawa}}\ and\ \bibinfo {author} {\bibfnamefont {B.}~\bibnamefont {Mera}},\
  }\bibfield  {title} {\bibinfo {title} {Relations between topology and the
  quantum metric for {{Chern}} insulators},\ }\href
  {https://doi.org/10.1103/PhysRevB.104.045103} {\bibfield  {journal} {\bibinfo
   {journal} {Phys. Rev. B}\ }\textbf {\bibinfo {volume} {104}},\ \bibinfo
  {pages} {045103} (\bibinfo {year} {2021})}\BibitemShut {NoStop}%
\bibitem [{\citenamefont {Mera}\ and\ \citenamefont
  {Ozawa}(2021)}]{Mera.Ozawa_PRB21_KahlerGeometryChern}%
  \BibitemOpen
  \bibfield  {author} {\bibinfo {author} {\bibfnamefont {B.}~\bibnamefont
  {Mera}}\ and\ \bibinfo {author} {\bibfnamefont {T.}~\bibnamefont {Ozawa}},\
  }\bibfield  {title} {\bibinfo {title} {K\"{a}hler geometry and chern
  insulators: {{Relations}} between topology and the quantum metric},\ }\href
  {https://doi.org/10.1103/PhysRevB.104.045104} {\bibfield  {journal} {\bibinfo
   {journal} {Phys. Rev. B}\ }\textbf {\bibinfo {volume} {104}},\ \bibinfo
  {pages} {045104} (\bibinfo {year} {2021})}\BibitemShut {NoStop}%
\bibitem [{\citenamefont {Ledwith}\ \emph {et~al.}(2023)\citenamefont
  {Ledwith}, \citenamefont {Vishwanath},\ and\ \citenamefont
  {Parker}}]{Ledwith.Parker_PRB23_VortexabilityUnifyingCriterion}%
  \BibitemOpen
  \bibfield  {author} {\bibinfo {author} {\bibfnamefont {P.~J.}\ \bibnamefont
  {Ledwith}}, \bibinfo {author} {\bibfnamefont {A.}~\bibnamefont
  {Vishwanath}},\ and\ \bibinfo {author} {\bibfnamefont {D.~E.}\ \bibnamefont
  {Parker}},\ }\bibfield  {title} {\bibinfo {title} {Vortexability: {{A}}
  unifying criterion for ideal fractional {{Chern}} insulators},\ }\href
  {https://doi.org/10.1103/PhysRevB.108.205144} {\bibfield  {journal} {\bibinfo
   {journal} {Phys. Rev. B}\ }\textbf {\bibinfo {volume} {108}},\ \bibinfo
  {pages} {205144} (\bibinfo {year} {2023})}\BibitemShut {NoStop}%
\bibitem [{\citenamefont {Estienne}\ \emph {et~al.}(2023)\citenamefont
  {Estienne}, \citenamefont {Regnault},\ and\ \citenamefont
  {Cr{\'e}pel}}]{Estienne.Crepel_PRR23_IdealChernBands}%
  \BibitemOpen
  \bibfield  {author} {\bibinfo {author} {\bibfnamefont {B.}~\bibnamefont
  {Estienne}}, \bibinfo {author} {\bibfnamefont {N.}~\bibnamefont {Regnault}},\
  and\ \bibinfo {author} {\bibfnamefont {V.}~\bibnamefont {Cr{\'e}pel}},\
  }\bibfield  {title} {\bibinfo {title} {Ideal {{Chern}} bands as {{Landau}}
  levels in curved space},\ }\href
  {https://doi.org/10.1103/PhysRevResearch.5.L032048} {\bibfield  {journal}
  {\bibinfo  {journal} {Phys. Rev. Res.}\ }\textbf {\bibinfo {volume} {5}},\
  \bibinfo {pages} {L032048} (\bibinfo {year} {2023})}\BibitemShut {NoStop}%
\bibitem [{\citenamefont {{Morales-Dur{\'a}n}}\ \emph
  {et~al.}(2024)\citenamefont {{Morales-Dur{\'a}n}}, \citenamefont {Wei},
  \citenamefont {Shi},\ and\ \citenamefont
  {MacDonald}}]{Morales-Duran.MacDonald_PRL24_MagicAnglesFractional}%
  \BibitemOpen
  \bibfield  {author} {\bibinfo {author} {\bibfnamefont {N.}~\bibnamefont
  {{Morales-Dur{\'a}n}}}, \bibinfo {author} {\bibfnamefont {N.}~\bibnamefont
  {Wei}}, \bibinfo {author} {\bibfnamefont {J.}~\bibnamefont {Shi}},\ and\
  \bibinfo {author} {\bibfnamefont {A.~H.}\ \bibnamefont {MacDonald}},\
  }\bibfield  {title} {\bibinfo {title} {Magic {{Angles}} and {{Fractional
  Chern Insulators}} in {{Twisted Homobilayer Transition Metal
  Dichalcogenides}}},\ }\href {https://doi.org/10.1103/PhysRevLett.132.096602}
  {\bibfield  {journal} {\bibinfo  {journal} {Phys. Rev. Lett.}\ }\textbf
  {\bibinfo {volume} {132}},\ \bibinfo {pages} {096602} (\bibinfo {year}
  {2024})}\BibitemShut {NoStop}%
\bibitem [{\citenamefont {Shi}\ \emph {et~al.}(2024)\citenamefont {Shi},
  \citenamefont {{Morales-Dur{\'a}n}}, \citenamefont {Khalaf},\ and\
  \citenamefont
  {MacDonald}}]{Shi.MacDonald_PRB24_AdiabaticApproximationAharonovCasher}%
  \BibitemOpen
  \bibfield  {author} {\bibinfo {author} {\bibfnamefont {J.}~\bibnamefont
  {Shi}}, \bibinfo {author} {\bibfnamefont {N.}~\bibnamefont
  {{Morales-Dur{\'a}n}}}, \bibinfo {author} {\bibfnamefont {E.}~\bibnamefont
  {Khalaf}},\ and\ \bibinfo {author} {\bibfnamefont {A.~H.}\ \bibnamefont
  {MacDonald}},\ }\bibfield  {title} {\bibinfo {title} {Adiabatic approximation
  and {{Aharonov-Casher}} bands in twisted homobilayer transition metal
  dichalcogenides},\ }\href {https://doi.org/10.1103/PhysRevB.110.035130}
  {\bibfield  {journal} {\bibinfo  {journal} {Phys. Rev. B}\ }\textbf {\bibinfo
  {volume} {110}},\ \bibinfo {pages} {035130} (\bibinfo {year}
  {2024})}\BibitemShut {NoStop}%
\bibitem [{\citenamefont {Li}\ and\ \citenamefont
  {Wu}(2025)}]{Li.Wu_PRB25_VariationalMappingChern}%
  \BibitemOpen
  \bibfield  {author} {\bibinfo {author} {\bibfnamefont {B.}~\bibnamefont
  {Li}}\ and\ \bibinfo {author} {\bibfnamefont {F.}~\bibnamefont {Wu}},\
  }\bibfield  {title} {\bibinfo {title} {Variational mapping of {{Chern}} bands
  to {{Landau}} levels: {{Application}} to fractional {{Chern}} insulators in
  twisted {$\rm MoTe_2$}},\ }\href
  {https://doi.org/10.1103/PhysRevB.111.125122} {\bibfield  {journal} {\bibinfo
   {journal} {Phys. Rev. B}\ }\textbf {\bibinfo {volume} {111}},\ \bibinfo
  {pages} {125122} (\bibinfo {year} {2025})}\BibitemShut {NoStop}%
\bibitem [{\citenamefont {Tan}\ and\ \citenamefont
  {Devakul}(2024)}]{Tan.Devakul_PRX24_ParentBerryCurvature}%
  \BibitemOpen
  \bibfield  {author} {\bibinfo {author} {\bibfnamefont {T.}~\bibnamefont
  {Tan}}\ and\ \bibinfo {author} {\bibfnamefont {T.}~\bibnamefont {Devakul}},\
  }\bibfield  {title} {\bibinfo {title} {Parent {{Berry Curvature}} and the
  {{Ideal Anomalous Hall Crystal}}},\ }\href
  {https://doi.org/10.1103/PhysRevX.14.041040} {\bibfield  {journal} {\bibinfo
  {journal} {Phys. Rev. X}\ }\textbf {\bibinfo {volume} {14}},\ \bibinfo
  {pages} {041040} (\bibinfo {year} {2024})}\BibitemShut {NoStop}%
\bibitem [{\citenamefont {Dong}\ \emph {et~al.}(2024)\citenamefont {Dong},
  \citenamefont {Wang}, \citenamefont {Wang}, \citenamefont {Soejima},
  \citenamefont {Zaletel}, \citenamefont {Vishwanath},\ and\ \citenamefont
  {Parker}}]{Dong.Parker_PRL24_AnomalousHallCrystals}%
  \BibitemOpen
  \bibfield  {author} {\bibinfo {author} {\bibfnamefont {J.}~\bibnamefont
  {Dong}}, \bibinfo {author} {\bibfnamefont {T.}~\bibnamefont {Wang}}, \bibinfo
  {author} {\bibfnamefont {T.}~\bibnamefont {Wang}}, \bibinfo {author}
  {\bibfnamefont {T.}~\bibnamefont {Soejima}}, \bibinfo {author} {\bibfnamefont
  {M.~P.}\ \bibnamefont {Zaletel}}, \bibinfo {author} {\bibfnamefont
  {A.}~\bibnamefont {Vishwanath}},\ and\ \bibinfo {author} {\bibfnamefont
  {D.~E.}\ \bibnamefont {Parker}},\ }\bibfield  {title} {\bibinfo {title}
  {Anomalous {{Hall Crystals}} in {{Rhombohedral Multilayer Graphene}}. {{I}}.
  {{Interaction-Driven Chern Bands}} and {{Fractional Quantum Hall States}} at
  {{Zero Magnetic Field}}},\ }\href
  {https://doi.org/10.1103/PhysRevLett.133.206503} {\bibfield  {journal}
  {\bibinfo  {journal} {Phys. Rev. Lett.}\ }\textbf {\bibinfo {volume} {133}},\
  \bibinfo {pages} {206503} (\bibinfo {year} {2024})}\BibitemShut {NoStop}%
\bibitem [{\citenamefont {Tan}\ \emph {et~al.}(2025)\citenamefont {Tan},
  \citenamefont {{May-Mann}},\ and\ \citenamefont
  {Devakul}}]{Tan.Devakul_PRL25_VariationalWaveFunctionAnalysis}%
  \BibitemOpen
  \bibfield  {author} {\bibinfo {author} {\bibfnamefont {T.}~\bibnamefont
  {Tan}}, \bibinfo {author} {\bibfnamefont {J.}~\bibnamefont {{May-Mann}}},\
  and\ \bibinfo {author} {\bibfnamefont {T.}~\bibnamefont {Devakul}},\
  }\bibfield  {title} {\bibinfo {title} {Variational {{Wave-Function Analysis}}
  of the {{Fractional Anomalous Hall Crystal}}},\ }\href
  {https://doi.org/10.1103/dd2d-kk3w} {\bibfield  {journal} {\bibinfo
  {journal} {Phys. Rev. Lett.}\ }\textbf {\bibinfo {volume} {135}},\ \bibinfo
  {pages} {036604} (\bibinfo {year} {2025})}\BibitemShut {NoStop}%
\bibitem [{\citenamefont {Cai}\ \emph {et~al.}(2023)\citenamefont {Cai},
  \citenamefont {Anderson}, \citenamefont {Wang}, \citenamefont {Zhang},
  \citenamefont {Liu}, \citenamefont {Holtzmann}, \citenamefont {Zhang},
  \citenamefont {Fan}, \citenamefont {Taniguchi}, \citenamefont {Watanabe},
  \citenamefont {Ran}, \citenamefont {Cao}, \citenamefont {Fu}, \citenamefont
  {Xiao}, \citenamefont {Yao},\ and\ \citenamefont
  {Xu}}]{Cai.Xu_N23_SignaturesFractionalQuantum}%
  \BibitemOpen
  \bibfield  {author} {\bibinfo {author} {\bibfnamefont {J.}~\bibnamefont
  {Cai}}, \bibinfo {author} {\bibfnamefont {E.}~\bibnamefont {Anderson}},
  \bibinfo {author} {\bibfnamefont {C.}~\bibnamefont {Wang}}, \bibinfo {author}
  {\bibfnamefont {X.}~\bibnamefont {Zhang}}, \bibinfo {author} {\bibfnamefont
  {X.}~\bibnamefont {Liu}}, \bibinfo {author} {\bibfnamefont {W.}~\bibnamefont
  {Holtzmann}}, \bibinfo {author} {\bibfnamefont {Y.}~\bibnamefont {Zhang}},
  \bibinfo {author} {\bibfnamefont {F.}~\bibnamefont {Fan}}, \bibinfo {author}
  {\bibfnamefont {T.}~\bibnamefont {Taniguchi}}, \bibinfo {author}
  {\bibfnamefont {K.}~\bibnamefont {Watanabe}}, \bibinfo {author}
  {\bibfnamefont {Y.}~\bibnamefont {Ran}}, \bibinfo {author} {\bibfnamefont
  {T.}~\bibnamefont {Cao}}, \bibinfo {author} {\bibfnamefont {L.}~\bibnamefont
  {Fu}}, \bibinfo {author} {\bibfnamefont {D.}~\bibnamefont {Xiao}}, \bibinfo
  {author} {\bibfnamefont {W.}~\bibnamefont {Yao}},\ and\ \bibinfo {author}
  {\bibfnamefont {X.}~\bibnamefont {Xu}},\ }\bibfield  {title} {\bibinfo
  {title} {Signatures of fractional quantum anomalous {{Hall}} states in
  twisted {$\mathrm{MoTe}_2$}},\ }\href
  {https://doi.org/10.1038/s41586-023-06289-w} {\bibfield  {journal} {\bibinfo
  {journal} {Nature}\ }\textbf {\bibinfo {volume} {622}},\ \bibinfo {pages}
  {63} (\bibinfo {year} {2023})}\BibitemShut {NoStop}%
\bibitem [{\citenamefont {Zeng}\ \emph {et~al.}(2023)\citenamefont {Zeng},
  \citenamefont {Xia}, \citenamefont {Kang}, \citenamefont {Zhu}, \citenamefont
  {Kn{\"u}ppel}, \citenamefont {Vaswani}, \citenamefont {Watanabe},
  \citenamefont {Taniguchi}, \citenamefont {Mak},\ and\ \citenamefont
  {Shan}}]{Zeng.Shan_N23_ThermodynamicEvidenceFractional}%
  \BibitemOpen
  \bibfield  {author} {\bibinfo {author} {\bibfnamefont {Y.}~\bibnamefont
  {Zeng}}, \bibinfo {author} {\bibfnamefont {Z.}~\bibnamefont {Xia}}, \bibinfo
  {author} {\bibfnamefont {K.}~\bibnamefont {Kang}}, \bibinfo {author}
  {\bibfnamefont {J.}~\bibnamefont {Zhu}}, \bibinfo {author} {\bibfnamefont
  {P.}~\bibnamefont {Kn{\"u}ppel}}, \bibinfo {author} {\bibfnamefont
  {C.}~\bibnamefont {Vaswani}}, \bibinfo {author} {\bibfnamefont
  {K.}~\bibnamefont {Watanabe}}, \bibinfo {author} {\bibfnamefont
  {T.}~\bibnamefont {Taniguchi}}, \bibinfo {author} {\bibfnamefont {K.~F.}\
  \bibnamefont {Mak}},\ and\ \bibinfo {author} {\bibfnamefont {J.}~\bibnamefont
  {Shan}},\ }\bibfield  {title} {\bibinfo {title} {Thermodynamic evidence of
  fractional {{Chern}} insulator in moir{\'e} {{MoTe2}}},\ }\href
  {https://doi.org/10.1038/s41586-023-06452-3} {\bibfield  {journal} {\bibinfo
  {journal} {Nature}\ }\textbf {\bibinfo {volume} {622}},\ \bibinfo {pages}
  {69} (\bibinfo {year} {2023})}\BibitemShut {NoStop}%
\bibitem [{\citenamefont {Park}\ \emph {et~al.}(2023)\citenamefont {Park},
  \citenamefont {Cai}, \citenamefont {Anderson}, \citenamefont {Zhang},
  \citenamefont {Zhu}, \citenamefont {Liu}, \citenamefont {Wang}, \citenamefont
  {Holtzmann}, \citenamefont {Hu}, \citenamefont {Liu}, \citenamefont
  {Taniguchi}, \citenamefont {Watanabe}, \citenamefont {Chu}, \citenamefont
  {Cao}, \citenamefont {Fu}, \citenamefont {Yao}, \citenamefont {Chang},
  \citenamefont {Cobden}, \citenamefont {Xiao},\ and\ \citenamefont
  {Xu}}]{Park.Xu_N23_ObservationFractionallyQuantized}%
  \BibitemOpen
  \bibfield  {author} {\bibinfo {author} {\bibfnamefont {H.}~\bibnamefont
  {Park}}, \bibinfo {author} {\bibfnamefont {J.}~\bibnamefont {Cai}}, \bibinfo
  {author} {\bibfnamefont {E.}~\bibnamefont {Anderson}}, \bibinfo {author}
  {\bibfnamefont {Y.}~\bibnamefont {Zhang}}, \bibinfo {author} {\bibfnamefont
  {J.}~\bibnamefont {Zhu}}, \bibinfo {author} {\bibfnamefont {X.}~\bibnamefont
  {Liu}}, \bibinfo {author} {\bibfnamefont {C.}~\bibnamefont {Wang}}, \bibinfo
  {author} {\bibfnamefont {W.}~\bibnamefont {Holtzmann}}, \bibinfo {author}
  {\bibfnamefont {C.}~\bibnamefont {Hu}}, \bibinfo {author} {\bibfnamefont
  {Z.}~\bibnamefont {Liu}}, \bibinfo {author} {\bibfnamefont {T.}~\bibnamefont
  {Taniguchi}}, \bibinfo {author} {\bibfnamefont {K.}~\bibnamefont {Watanabe}},
  \bibinfo {author} {\bibfnamefont {J.-H.}\ \bibnamefont {Chu}}, \bibinfo
  {author} {\bibfnamefont {T.}~\bibnamefont {Cao}}, \bibinfo {author}
  {\bibfnamefont {L.}~\bibnamefont {Fu}}, \bibinfo {author} {\bibfnamefont
  {W.}~\bibnamefont {Yao}}, \bibinfo {author} {\bibfnamefont {C.-Z.}\
  \bibnamefont {Chang}}, \bibinfo {author} {\bibfnamefont {D.}~\bibnamefont
  {Cobden}}, \bibinfo {author} {\bibfnamefont {D.}~\bibnamefont {Xiao}},\ and\
  \bibinfo {author} {\bibfnamefont {X.}~\bibnamefont {Xu}},\ }\bibfield
  {title} {\bibinfo {title} {Observation of fractionally quantized anomalous
  {{Hall}} effect},\ }\href {https://doi.org/10.1038/s41586-023-06536-0}
  {\bibfield  {journal} {\bibinfo  {journal} {Nature}\ }\textbf {\bibinfo
  {volume} {622}},\ \bibinfo {pages} {74} (\bibinfo {year} {2023})}\BibitemShut
  {NoStop}%
\bibitem [{\citenamefont {Xu}\ \emph {et~al.}(2023)\citenamefont {Xu},
  \citenamefont {Sun}, \citenamefont {Jia}, \citenamefont {Liu}, \citenamefont
  {Xu}, \citenamefont {Li}, \citenamefont {Gu}, \citenamefont {Watanabe},
  \citenamefont {Taniguchi}, \citenamefont {Tong}, \citenamefont {Jia},
  \citenamefont {Shi}, \citenamefont {Jiang}, \citenamefont {Zhang},
  \citenamefont {Liu},\ and\ \citenamefont
  {Li}}]{Xu.Li_PRX23_ObservationIntegerFractional}%
  \BibitemOpen
  \bibfield  {author} {\bibinfo {author} {\bibfnamefont {F.}~\bibnamefont
  {Xu}}, \bibinfo {author} {\bibfnamefont {Z.}~\bibnamefont {Sun}}, \bibinfo
  {author} {\bibfnamefont {T.}~\bibnamefont {Jia}}, \bibinfo {author}
  {\bibfnamefont {C.}~\bibnamefont {Liu}}, \bibinfo {author} {\bibfnamefont
  {C.}~\bibnamefont {Xu}}, \bibinfo {author} {\bibfnamefont {C.}~\bibnamefont
  {Li}}, \bibinfo {author} {\bibfnamefont {Y.}~\bibnamefont {Gu}}, \bibinfo
  {author} {\bibfnamefont {K.}~\bibnamefont {Watanabe}}, \bibinfo {author}
  {\bibfnamefont {T.}~\bibnamefont {Taniguchi}}, \bibinfo {author}
  {\bibfnamefont {B.}~\bibnamefont {Tong}}, \bibinfo {author} {\bibfnamefont
  {J.}~\bibnamefont {Jia}}, \bibinfo {author} {\bibfnamefont {Z.}~\bibnamefont
  {Shi}}, \bibinfo {author} {\bibfnamefont {S.}~\bibnamefont {Jiang}}, \bibinfo
  {author} {\bibfnamefont {Y.}~\bibnamefont {Zhang}}, \bibinfo {author}
  {\bibfnamefont {X.}~\bibnamefont {Liu}},\ and\ \bibinfo {author}
  {\bibfnamefont {T.}~\bibnamefont {Li}},\ }\bibfield  {title} {\bibinfo
  {title} {Observation of {{Integer}} and {{Fractional Quantum Anomalous Hall
  Effects}} in {{Twisted Bilayer}} {$\mathrm{MoTe}_2$}},\ }\href
  {https://doi.org/10.1103/PhysRevX.13.031037} {\bibfield  {journal} {\bibinfo
  {journal} {Phys. Rev. X}\ }\textbf {\bibinfo {volume} {13}},\ \bibinfo
  {pages} {031037} (\bibinfo {year} {2023})}\BibitemShut {NoStop}%
\bibitem [{\citenamefont {Lu}\ \emph {et~al.}(2024)\citenamefont {Lu},
  \citenamefont {Han}, \citenamefont {Yao}, \citenamefont {Reddy},
  \citenamefont {Yang}, \citenamefont {Seo}, \citenamefont {Watanabe},
  \citenamefont {Taniguchi}, \citenamefont {Fu},\ and\ \citenamefont
  {Ju}}]{Lu.Ju_N24_FractionalQuantumAnomalous}%
  \BibitemOpen
  \bibfield  {author} {\bibinfo {author} {\bibfnamefont {Z.}~\bibnamefont
  {Lu}}, \bibinfo {author} {\bibfnamefont {T.}~\bibnamefont {Han}}, \bibinfo
  {author} {\bibfnamefont {Y.}~\bibnamefont {Yao}}, \bibinfo {author}
  {\bibfnamefont {A.~P.}\ \bibnamefont {Reddy}}, \bibinfo {author}
  {\bibfnamefont {J.}~\bibnamefont {Yang}}, \bibinfo {author} {\bibfnamefont
  {J.}~\bibnamefont {Seo}}, \bibinfo {author} {\bibfnamefont {K.}~\bibnamefont
  {Watanabe}}, \bibinfo {author} {\bibfnamefont {T.}~\bibnamefont {Taniguchi}},
  \bibinfo {author} {\bibfnamefont {L.}~\bibnamefont {Fu}},\ and\ \bibinfo
  {author} {\bibfnamefont {L.}~\bibnamefont {Ju}},\ }\bibfield  {title}
  {\bibinfo {title} {Fractional quantum anomalous {{Hall}} effect in multilayer
  graphene},\ }\href {https://doi.org/10.1038/s41586-023-07010-7} {\bibfield
  {journal} {\bibinfo  {journal} {Nature}\ }\textbf {\bibinfo {volume} {626}},\
  \bibinfo {pages} {759} (\bibinfo {year} {2024})}\BibitemShut {NoStop}%
\bibitem [{Note1()}]{Note1}%
  \BibitemOpen
  \bibinfo {note} {In general this field is a function that parametrizes
  single-particle wavefunctions and not necessarily a physical
  field.}\BibitemShut {Stop}%
\bibitem [{\citenamefont
  {Laughlin}(1983)}]{Laughlin_PRL83_AnomalousQuantumHall}%
  \BibitemOpen
  \bibfield  {author} {\bibinfo {author} {\bibfnamefont {R.~B.}\ \bibnamefont
  {Laughlin}},\ }\bibfield  {title} {\bibinfo {title} {Anomalous {{Quantum Hall
  Effect}}: {{An Incompressible Quantum Fluid}} with {{Fractionally Charged
  Excitations}}},\ }\href {https://doi.org/10.1103/PhysRevLett.50.1395}
  {\bibfield  {journal} {\bibinfo  {journal} {Phys. Rev. Lett.}\ }\textbf
  {\bibinfo {volume} {50}},\ \bibinfo {pages} {1395} (\bibinfo {year}
  {1983})}\BibitemShut {NoStop}%
\bibitem [{\citenamefont {Wolf}\ \emph {et~al.}(2025)\citenamefont {Wolf},
  \citenamefont {Chao}, \citenamefont {MacDonald},\ and\ \citenamefont
  {Su}}]{Wolf.Su_PRL25_IntrabandCollectiveExcitations}%
  \BibitemOpen
  \bibfield  {author} {\bibinfo {author} {\bibfnamefont {T.}~\bibnamefont
  {Wolf}}, \bibinfo {author} {\bibfnamefont {Y.-C.}\ \bibnamefont {Chao}},
  \bibinfo {author} {\bibfnamefont {A.~H.}\ \bibnamefont {MacDonald}},\ and\
  \bibinfo {author} {\bibfnamefont {J.-J.}\ \bibnamefont {Su}},\ }\bibfield
  {title} {\bibinfo {title} {Intraband {{Collective Excitations}} and {{Spatial
  Correlations}} in {{Fractional Chern Insulators}}},\ }\href
  {https://doi.org/10.1103/PhysRevLett.134.116501} {\bibfield  {journal}
  {\bibinfo  {journal} {Phys. Rev. Lett.}\ }\textbf {\bibinfo {volume} {134}},\
  \bibinfo {pages} {116501} (\bibinfo {year} {2025})}\BibitemShut {NoStop}%
\bibitem [{\citenamefont {Clerouin}\ \emph {et~al.}(1987)\citenamefont
  {Clerouin}, \citenamefont {Hansen},\ and\ \citenamefont
  {Piller}}]{Clerouin.Piller_PRA87_TwodimensionalClassicalElectron}%
  \BibitemOpen
  \bibfield  {author} {\bibinfo {author} {\bibfnamefont {J.}~\bibnamefont
  {Clerouin}}, \bibinfo {author} {\bibfnamefont {J.-P.}\ \bibnamefont
  {Hansen}},\ and\ \bibinfo {author} {\bibfnamefont {B.}~\bibnamefont
  {Piller}},\ }\bibfield  {title} {\bibinfo {title} {Two-dimensional classical
  electron gas in a periodic field: {{Delocalization}} and dielectric-plasma
  transition},\ }\href {https://doi.org/10.1103/PhysRevA.36.2793} {\bibfield
  {journal} {\bibinfo  {journal} {Phys. Rev. A}\ }\textbf {\bibinfo {volume}
  {36}},\ \bibinfo {pages} {2793} (\bibinfo {year} {1987})}\BibitemShut
  {NoStop}%
\bibitem [{\citenamefont {Alastuey}\ \emph {et~al.}(1988)\citenamefont
  {Alastuey}, \citenamefont {Cornu},\ and\ \citenamefont
  {Jancovici}}]{Alastuey.Jancovici_PRA88_CommentTwodimensionalClassical}%
  \BibitemOpen
  \bibfield  {author} {\bibinfo {author} {\bibfnamefont {A.}~\bibnamefont
  {Alastuey}}, \bibinfo {author} {\bibfnamefont {F.}~\bibnamefont {Cornu}},\
  and\ \bibinfo {author} {\bibfnamefont {B.}~\bibnamefont {Jancovici}},\
  }\bibfield  {title} {\bibinfo {title} {Comment on ``{{Two-dimensional}}
  classical electron gas in a periodic field: {{Delocalization}} and
  dielectric-plasma transition''},\ }\href
  {https://doi.org/10.1103/PhysRevA.38.4916} {\bibfield  {journal} {\bibinfo
  {journal} {Phys. Rev. A}\ }\textbf {\bibinfo {volume} {38}},\ \bibinfo
  {pages} {4916} (\bibinfo {year} {1988})}\BibitemShut {NoStop}%
\bibitem [{\citenamefont {Choquard}\ \emph {et~al.}(1989)\citenamefont
  {Choquard}, \citenamefont {Piller}, \citenamefont {Rentsch}, \citenamefont
  {Cl{\'e}rouin},\ and\ \citenamefont
  {Hansen}}]{Choquard.Hansen_PRA89_IonizationPhaseDiagram}%
  \BibitemOpen
  \bibfield  {author} {\bibinfo {author} {\bibfnamefont {P.}~\bibnamefont
  {Choquard}}, \bibinfo {author} {\bibfnamefont {B.}~\bibnamefont {Piller}},
  \bibinfo {author} {\bibfnamefont {R.}~\bibnamefont {Rentsch}}, \bibinfo
  {author} {\bibfnamefont {J.}~\bibnamefont {Cl{\'e}rouin}},\ and\ \bibinfo
  {author} {\bibfnamefont {J.-P.}\ \bibnamefont {Hansen}},\ }\bibfield  {title}
  {\bibinfo {title} {Ionization and phase diagram of classical {{Thomson}}
  atoms on a triangular lattice},\ }\href
  {https://doi.org/10.1103/PhysRevA.40.931} {\bibfield  {journal} {\bibinfo
  {journal} {Phys. Rev. A}\ }\textbf {\bibinfo {volume} {40}},\ \bibinfo
  {pages} {931} (\bibinfo {year} {1989})}\BibitemShut {NoStop}%
\bibitem [{\citenamefont
  {Haldane}(1983)}]{Haldane_PRL83_FractionalQuantizationHall}%
  \BibitemOpen
  \bibfield  {author} {\bibinfo {author} {\bibfnamefont {F.~D.~M.}\
  \bibnamefont {Haldane}},\ }\bibfield  {title} {\bibinfo {title} {Fractional
  {{Quantization}} of the {{Hall Effect}}: {{A Hierarchy}} of {{Incompressible
  Quantum Fluid States}}},\ }\href {https://doi.org/10.1103/PhysRevLett.51.605}
  {\bibfield  {journal} {\bibinfo  {journal} {Phys. Rev. Lett.}\ }\textbf
  {\bibinfo {volume} {51}},\ \bibinfo {pages} {605} (\bibinfo {year}
  {1983})}\BibitemShut {NoStop}%
\bibitem [{\citenamefont {Trugman}\ and\ \citenamefont
  {Kivelson}(1985)}]{Trugman.Kivelson_PRB85_ExactResultsFractional}%
  \BibitemOpen
  \bibfield  {author} {\bibinfo {author} {\bibfnamefont {S.~A.}\ \bibnamefont
  {Trugman}}\ and\ \bibinfo {author} {\bibfnamefont {S.}~\bibnamefont
  {Kivelson}},\ }\bibfield  {title} {\bibinfo {title} {Exact results for the
  fractional quantum {{Hall}} effect with general interactions},\ }\href
  {https://doi.org/10.1103/PhysRevB.31.5280} {\bibfield  {journal} {\bibinfo
  {journal} {Phys. Rev. B}\ }\textbf {\bibinfo {volume} {31}},\ \bibinfo
  {pages} {5280} (\bibinfo {year} {1985})}\BibitemShut {NoStop}%
\bibitem [{\citenamefont {Caillol}\ \emph {et~al.}(1982)\citenamefont
  {Caillol}, \citenamefont {Levesque}, \citenamefont {Weis},\ and\
  \citenamefont {Hansen}}]{Caillol.Hansen_JSP82_MonteCarloStudy}%
  \BibitemOpen
  \bibfield  {author} {\bibinfo {author} {\bibfnamefont {J.~M.}\ \bibnamefont
  {Caillol}}, \bibinfo {author} {\bibfnamefont {D.}~\bibnamefont {Levesque}},
  \bibinfo {author} {\bibfnamefont {J.~J.}\ \bibnamefont {Weis}},\ and\
  \bibinfo {author} {\bibfnamefont {J.~P.}\ \bibnamefont {Hansen}},\ }\bibfield
   {title} {\bibinfo {title} {A {{Monte Carlo}} study of the classical
  two-dimensional one-component plasma},\ }\href
  {https://doi.org/10.1007/BF01012609} {\bibfield  {journal} {\bibinfo
  {journal} {J Stat Phys}\ }\textbf {\bibinfo {volume} {28}},\ \bibinfo {pages}
  {325} (\bibinfo {year} {1982})}\BibitemShut {NoStop}%
\bibitem [{\citenamefont {Herland}\ \emph {et~al.}(2013)\citenamefont
  {Herland}, \citenamefont {Babaev}, \citenamefont {Bonderson}, \citenamefont
  {Gurarie}, \citenamefont {Nayak}, \citenamefont {Radzihovsky},\ and\
  \citenamefont
  {Sudb{\o}}}]{Herland.Sudbo_PRB13_FreezingUnconventionalTwodimensional}%
  \BibitemOpen
  \bibfield  {author} {\bibinfo {author} {\bibfnamefont {E.~V.}\ \bibnamefont
  {Herland}}, \bibinfo {author} {\bibfnamefont {E.}~\bibnamefont {Babaev}},
  \bibinfo {author} {\bibfnamefont {P.}~\bibnamefont {Bonderson}}, \bibinfo
  {author} {\bibfnamefont {V.}~\bibnamefont {Gurarie}}, \bibinfo {author}
  {\bibfnamefont {C.}~\bibnamefont {Nayak}}, \bibinfo {author} {\bibfnamefont
  {L.}~\bibnamefont {Radzihovsky}},\ and\ \bibinfo {author} {\bibfnamefont
  {A.}~\bibnamefont {Sudb{\o}}},\ }\bibfield  {title} {\bibinfo {title}
  {Freezing of an unconventional two-dimensional plasma},\ }\href
  {https://doi.org/10.1103/PhysRevB.87.075117} {\bibfield  {journal} {\bibinfo
  {journal} {Phys. Rev. B}\ }\textbf {\bibinfo {volume} {87}},\ \bibinfo
  {pages} {075117} (\bibinfo {year} {2013})}\BibitemShut {NoStop}%
\bibitem [{\citenamefont {Wu}\ \emph {et~al.}(2019)\citenamefont {Wu},
  \citenamefont {Lovorn}, \citenamefont {Tutuc}, \citenamefont {Martin},\ and\
  \citenamefont {MacDonald}}]{Wu.MacDonald_PRL19_TopologicalInsulatorsTwisted}%
  \BibitemOpen
  \bibfield  {author} {\bibinfo {author} {\bibfnamefont {F.}~\bibnamefont
  {Wu}}, \bibinfo {author} {\bibfnamefont {T.}~\bibnamefont {Lovorn}}, \bibinfo
  {author} {\bibfnamefont {E.}~\bibnamefont {Tutuc}}, \bibinfo {author}
  {\bibfnamefont {I.}~\bibnamefont {Martin}},\ and\ \bibinfo {author}
  {\bibfnamefont {A.~H.}\ \bibnamefont {MacDonald}},\ }\bibfield  {title}
  {\bibinfo {title} {Topological {{Insulators}} in {{Twisted Transition Metal
  Dichalcogenide Homobilayers}}},\ }\href
  {https://doi.org/10.1103/PhysRevLett.122.086402} {\bibfield  {journal}
  {\bibinfo  {journal} {Phys. Rev. Lett.}\ }\textbf {\bibinfo {volume} {122}},\
  \bibinfo {pages} {086402} (\bibinfo {year} {2019})}\BibitemShut {NoStop}%
\bibitem [{\citenamefont {Martin}\ and\ \citenamefont
  {Gruber}(1983)}]{Martin.Gruber_JSP83_NewProofStillingerLovett}%
  \BibitemOpen
  \bibfield  {author} {\bibinfo {author} {\bibfnamefont {{\relax Ph}.~A.}\
  \bibnamefont {Martin}}\ and\ \bibinfo {author} {\bibfnamefont {{\relax
  Ch}.}~\bibnamefont {Gruber}},\ }\bibfield  {title} {\bibinfo {title} {A new
  proof of the {{Stillinger-Lovett}} complete shielding condition},\ }\href
  {https://doi.org/10.1007/BF01019506} {\bibfield  {journal} {\bibinfo
  {journal} {J Stat Phys}\ }\textbf {\bibinfo {volume} {31}},\ \bibinfo {pages}
  {691} (\bibinfo {year} {1983})}\BibitemShut {NoStop}%
\bibitem [{sup()}]{suppl}%
  \BibitemOpen
  \href@noop {} {}\bibinfo {note} {See Supplemental Material at
  URL-to-be-inserted for (1) linear response relation between the dielectric
  function and charge correlation function, (2) sum rule for the dielectric
  constant and its relation to particle position fluctuations, and (3)
  asymptotics of the cell averaged correlator at leading order of cluster
  expansion.}\BibitemShut {Stop}%
\bibitem [{\citenamefont {Mitchell}\ \emph {et~al.}(1977)\citenamefont
  {Mitchell}, \citenamefont {McQuarrie}, \citenamefont {Szabo},\ and\
  \citenamefont
  {Groeneveld}}]{Mitchell.Groeneveld_JSP77_SecondmomentConditionStillinger}%
  \BibitemOpen
  \bibfield  {author} {\bibinfo {author} {\bibfnamefont {D.~J.}\ \bibnamefont
  {Mitchell}}, \bibinfo {author} {\bibfnamefont {D.~A.}\ \bibnamefont
  {McQuarrie}}, \bibinfo {author} {\bibfnamefont {A.}~\bibnamefont {Szabo}},\
  and\ \bibinfo {author} {\bibfnamefont {J.}~\bibnamefont {Groeneveld}},\
  }\bibfield  {title} {\bibinfo {title} {On the second-moment condition of
  {{Stillinger}} and {{Lovett}}},\ }\href {https://doi.org/10.1007/BF01089374}
  {\bibfield  {journal} {\bibinfo  {journal} {J Stat Phys}\ }\textbf {\bibinfo
  {volume} {17}},\ \bibinfo {pages} {15} (\bibinfo {year} {1977})}\BibitemShut
  {NoStop}%
\bibitem [{\citenamefont {Martin}(1988)}]{Martin_RMP88_SumRulesCharged}%
  \BibitemOpen
  \bibfield  {author} {\bibinfo {author} {\bibfnamefont {{\relax Ph}.~A.}\
  \bibnamefont {Martin}},\ }\bibfield  {title} {\bibinfo {title} {Sum rules in
  charged fluids},\ }\href {https://doi.org/10.1103/RevModPhys.60.1075}
  {\bibfield  {journal} {\bibinfo  {journal} {Rev. Mod. Phys.}\ }\textbf
  {\bibinfo {volume} {60}},\ \bibinfo {pages} {1075} (\bibinfo {year}
  {1988})}\BibitemShut {NoStop}%
\bibitem [{\citenamefont {Wen}(2007)}]{Wen_07_QuantumFieldTheory}%
  \BibitemOpen
  \bibfield  {author} {\bibinfo {author} {\bibfnamefont {X.-G.}\ \bibnamefont
  {Wen}},\ }\href {https://doi.org/10.1093/acprof:oso/9780199227259.001.0001}
  {\emph {\bibinfo {title} {Quantum {{Field Theory}} of {{Many-Body Systems}}:
  {{From}} the {{Origin}} of {{Sound}} to an {{Origin}} of {{Light}} and
  {{Electrons}}}}}\ (\bibinfo  {publisher} {Oxford University Press},\ \bibinfo
  {year} {2007})\BibitemShut {NoStop}%
\bibitem [{\citenamefont {Hastings}\ and\ \citenamefont
  {Koma}(2006)}]{Hastings.Koma_CMP06_SpectralGapExponential}%
  \BibitemOpen
  \bibfield  {author} {\bibinfo {author} {\bibfnamefont {M.~B.}\ \bibnamefont
  {Hastings}}\ and\ \bibinfo {author} {\bibfnamefont {T.}~\bibnamefont
  {Koma}},\ }\bibfield  {title} {\bibinfo {title} {Spectral {{Gap}} and
  {{Exponential Decay}} of {{Correlations}}},\ }\href
  {https://doi.org/10.1007/s00220-006-0030-4} {\bibfield  {journal} {\bibinfo
  {journal} {Commun. Math. Phys.}\ }\textbf {\bibinfo {volume} {265}},\
  \bibinfo {pages} {781} (\bibinfo {year} {2006})}\BibitemShut {NoStop}%
\bibitem [{\citenamefont {Berezinski{\v
  i}}(1971)}]{Berezinskii_SPJ71_DestructionLongrangeOrder}%
  \BibitemOpen
  \bibfield  {author} {\bibinfo {author} {\bibfnamefont {V.~L.}\ \bibnamefont
  {Berezinski{\v i}}},\ }\bibfield  {title} {\bibinfo {title} {Destruction of
  {{Long-range Order}} in {{One-dimensional}} and {{Two-dimensional Systems}}
  having a {{Continuous Symmetry Group I}}. {{Classical Systems}}},\ }\href
  {http://jetp.ras.ru/cgi-bin/e/index/e/32/3/p493?a=list} {\bibfield  {journal}
  {\bibinfo  {journal} {Sov. Phys. JETP}\ }\textbf {\bibinfo {volume} {32}},\
  \bibinfo {pages} {493} (\bibinfo {year} {1971})}\BibitemShut {NoStop}%
\bibitem [{\citenamefont {Berezinski{\v
  i}}(1972)}]{Berezinskii_SPJ72_DestructionLongrangeOrder}%
  \BibitemOpen
  \bibfield  {author} {\bibinfo {author} {\bibfnamefont {V.~L.}\ \bibnamefont
  {Berezinski{\v i}}},\ }\bibfield  {title} {\bibinfo {title} {Destruction of
  {{Long-range Order}} in {{One-dimensional}} and {{Two-dimensional Systems
  Possessing}} a {{Continuous Symmetry Group}}. {{II}}. {{Quantum Systems}}},\
  }\href {http://jetp.ras.ru/cgi-bin/e/index/e/34/3/p610?a=list} {\bibfield
  {journal} {\bibinfo  {journal} {Sov. Phys. JETP}\ }\textbf {\bibinfo {volume}
  {34}},\ \bibinfo {pages} {610} (\bibinfo {year} {1972})}\BibitemShut
  {NoStop}%
\bibitem [{\citenamefont {Kosterlitz}\ and\ \citenamefont
  {Thouless}(1973)}]{Kosterlitz.Thouless_JPCSSP73_OrderingMetastabilityPhase}%
  \BibitemOpen
  \bibfield  {author} {\bibinfo {author} {\bibfnamefont {J.~M.}\ \bibnamefont
  {Kosterlitz}}\ and\ \bibinfo {author} {\bibfnamefont {D.~J.}\ \bibnamefont
  {Thouless}},\ }\bibfield  {title} {\bibinfo {title} {Ordering, metastability
  and phase transitions in two-dimensional systems},\ }\href
  {https://doi.org/10.1088/0022-3719/6/7/010} {\bibfield  {journal} {\bibinfo
  {journal} {J. Phys. C: Solid State Phys.}\ }\textbf {\bibinfo {volume} {6}},\
  \bibinfo {pages} {1181} (\bibinfo {year} {1973})}\BibitemShut {NoStop}%
\bibitem [{\citenamefont {Alastuey}\ and\ \citenamefont
  {Cornu}(1992)}]{Alastuey.Cornu_JSP92_CorrelationsKosterlitzThoulessPhase}%
  \BibitemOpen
  \bibfield  {author} {\bibinfo {author} {\bibfnamefont {A.}~\bibnamefont
  {Alastuey}}\ and\ \bibinfo {author} {\bibfnamefont {F.}~\bibnamefont
  {Cornu}},\ }\bibfield  {title} {\bibinfo {title} {Correlations in the
  {{Kosterlitz-Thouless}} phase of the two-dimensional {{Coulomb}} gas},\
  }\href {https://doi.org/10.1007/BF01060065} {\bibfield  {journal} {\bibinfo
  {journal} {J Stat Phys}\ }\textbf {\bibinfo {volume} {66}},\ \bibinfo {pages}
  {165} (\bibinfo {year} {1992})}\BibitemShut {NoStop}%
\bibitem [{\citenamefont {Alastuey}\ and\ \citenamefont
  {Forrester}(1995)}]{Alastuey.Forrester_JSP95_CorrelationsTwocomponentLoggas}%
  \BibitemOpen
  \bibfield  {author} {\bibinfo {author} {\bibfnamefont {A.}~\bibnamefont
  {Alastuey}}\ and\ \bibinfo {author} {\bibfnamefont {P.~J.}\ \bibnamefont
  {Forrester}},\ }\bibfield  {title} {\bibinfo {title} {Correlations in
  two-component log-gas systems},\ }\href {https://doi.org/10.1007/BF02179249}
  {\bibfield  {journal} {\bibinfo  {journal} {J Stat Phys}\ }\textbf {\bibinfo
  {volume} {81}},\ \bibinfo {pages} {579} (\bibinfo {year} {1995})}\BibitemShut
  {NoStop}%
\bibitem [{\citenamefont {Minnhagen}(1987)}]{minnhagen1987two}%
  \BibitemOpen
  \bibfield  {author} {\bibinfo {author} {\bibfnamefont {P.}~\bibnamefont
  {Minnhagen}},\ }\bibfield  {title} {\bibinfo {title} {The two-dimensional
  {{Coulomb}} gas, vortex unbinding, and superfluid-superconducting films},\
  }\href {https://doi.org/10.1103/RevModPhys.59.1001} {\bibfield  {journal}
  {\bibinfo  {journal} {Rev. Mod. Phys.}\ }\textbf {\bibinfo {volume} {59}},\
  \bibinfo {pages} {1001} (\bibinfo {year} {1987})}\BibitemShut {NoStop}%
\bibitem [{\citenamefont {Rana}\ and\ \citenamefont
  {Girvin}(1993)}]{rana1993soluble}%
  \BibitemOpen
  \bibfield  {author} {\bibinfo {author} {\bibfnamefont {A.~E.}\ \bibnamefont
  {Rana}}\ and\ \bibinfo {author} {\bibfnamefont {S.~M.}\ \bibnamefont
  {Girvin}},\ }\bibfield  {title} {\bibinfo {title} {Soluble supersymmetric
  quantum {{{\emph{XY}}}} model},\ }\href
  {https://doi.org/10.1103/PhysRevB.48.360} {\bibfield  {journal} {\bibinfo
  {journal} {Phys. Rev. B}\ }\textbf {\bibinfo {volume} {48}},\ \bibinfo
  {pages} {360} (\bibinfo {year} {1993})}\BibitemShut {NoStop}%
\bibitem [{\citenamefont {Ardonne}\ \emph {et~al.}(2004)\citenamefont
  {Ardonne}, \citenamefont {Fendley},\ and\ \citenamefont
  {Fradkin}}]{Ardonne.Fradkin_AoP04_TopologicalOrderConformal}%
  \BibitemOpen
  \bibfield  {author} {\bibinfo {author} {\bibfnamefont {E.}~\bibnamefont
  {Ardonne}}, \bibinfo {author} {\bibfnamefont {P.}~\bibnamefont {Fendley}},\
  and\ \bibinfo {author} {\bibfnamefont {E.}~\bibnamefont {Fradkin}},\
  }\bibfield  {title} {\bibinfo {title} {Topological order and conformal
  quantum critical points},\ }\href {https://doi.org/10.1016/j.aop.2004.01.004}
  {\bibfield  {journal} {\bibinfo  {journal} {Annals of Physics}\ }\textbf
  {\bibinfo {volume} {310}},\ \bibinfo {pages} {493} (\bibinfo {year}
  {2004})}\BibitemShut {NoStop}%
\bibitem [{\citenamefont {Fradkin}\ \emph {et~al.}(2004)\citenamefont
  {Fradkin}, \citenamefont {Huse}, \citenamefont {Moessner}, \citenamefont
  {Oganesyan},\ and\ \citenamefont
  {Sondhi}}]{Fradkin.Sondhi_PRB04_BipartiteRokhsarKivelsonPoints}%
  \BibitemOpen
  \bibfield  {author} {\bibinfo {author} {\bibfnamefont {E.}~\bibnamefont
  {Fradkin}}, \bibinfo {author} {\bibfnamefont {D.~A.}\ \bibnamefont {Huse}},
  \bibinfo {author} {\bibfnamefont {R.}~\bibnamefont {Moessner}}, \bibinfo
  {author} {\bibfnamefont {V.}~\bibnamefont {Oganesyan}},\ and\ \bibinfo
  {author} {\bibfnamefont {S.~L.}\ \bibnamefont {Sondhi}},\ }\bibfield  {title}
  {\bibinfo {title} {Bipartite {{Rokhsar--Kivelson}} points and {{Cantor}}
  deconfinement},\ }\href {https://doi.org/10.1103/PhysRevB.69.224415}
  {\bibfield  {journal} {\bibinfo  {journal} {Phys. Rev. B}\ }\textbf {\bibinfo
  {volume} {69}},\ \bibinfo {pages} {224415} (\bibinfo {year}
  {2004})}\BibitemShut {NoStop}%
\bibitem [{\citenamefont {Vishwanath}\ \emph {et~al.}(2004)\citenamefont
  {Vishwanath}, \citenamefont {Balents},\ and\ \citenamefont
  {Senthil}}]{Vishwanath.Senthil_PRB04_QuantumCriticalityDeconfinement}%
  \BibitemOpen
  \bibfield  {author} {\bibinfo {author} {\bibfnamefont {A.}~\bibnamefont
  {Vishwanath}}, \bibinfo {author} {\bibfnamefont {L.}~\bibnamefont
  {Balents}},\ and\ \bibinfo {author} {\bibfnamefont {T.}~\bibnamefont
  {Senthil}},\ }\bibfield  {title} {\bibinfo {title} {Quantum criticality and
  deconfinement in phase transitions between valence bond solids},\ }\href
  {https://doi.org/10.1103/PhysRevB.69.224416} {\bibfield  {journal} {\bibinfo
  {journal} {Phys. Rev. B}\ }\textbf {\bibinfo {volume} {69}},\ \bibinfo
  {pages} {224416} (\bibinfo {year} {2004})}\BibitemShut {NoStop}%
\bibitem [{\citenamefont {Isakov}\ \emph {et~al.}(2011)\citenamefont {Isakov},
  \citenamefont {Fendley}, \citenamefont {Ludwig}, \citenamefont {Trebst},\
  and\ \citenamefont {Troyer}}]{isakov2011dynamics}%
  \BibitemOpen
  \bibfield  {author} {\bibinfo {author} {\bibfnamefont {S.~V.}\ \bibnamefont
  {Isakov}}, \bibinfo {author} {\bibfnamefont {P.}~\bibnamefont {Fendley}},
  \bibinfo {author} {\bibfnamefont {A.~W.~W.}\ \bibnamefont {Ludwig}}, \bibinfo
  {author} {\bibfnamefont {S.}~\bibnamefont {Trebst}},\ and\ \bibinfo {author}
  {\bibfnamefont {M.}~\bibnamefont {Troyer}},\ }\bibfield  {title} {\bibinfo
  {title} {Dynamics at and near conformal quantum critical points},\ }\href
  {https://doi.org/10.1103/PhysRevB.83.125114} {\bibfield  {journal} {\bibinfo
  {journal} {Phys. Rev. B}\ }\textbf {\bibinfo {volume} {83}},\ \bibinfo
  {pages} {125114} (\bibinfo {year} {2011})}\BibitemShut {NoStop}%
\bibitem [{\citenamefont {Hsu}\ and\ \citenamefont
  {Fradkin}(2013)}]{Hsu.Fradkin_PRB13_DynamicalStabilityQuantum}%
  \BibitemOpen
  \bibfield  {author} {\bibinfo {author} {\bibfnamefont {B.}~\bibnamefont
  {Hsu}}\ and\ \bibinfo {author} {\bibfnamefont {E.}~\bibnamefont {Fradkin}},\
  }\bibfield  {title} {\bibinfo {title} {Dynamical stability of the quantum
  {{Lifshitz}} theory in 2+1 dimensions},\ }\href
  {https://doi.org/10.1103/PhysRevB.87.085102} {\bibfield  {journal} {\bibinfo
  {journal} {Phys. Rev. B}\ }\textbf {\bibinfo {volume} {87}},\ \bibinfo
  {pages} {085102} (\bibinfo {year} {2013})}\BibitemShut {NoStop}%
\end{thebibliography}%



\end{document}


\title{
--- Supplemental Material --- \\*[1em]
Instability of Laughlin FQH liquids into gapless power-law correlated states with continuous exponents in ideal Chern bands: rigorous results from plasma mapping
}

\author{Saranyo Moitra\,\orcidlink{0000-0001-7912-1961}}
\email{saranyo.moitra@uni-leipzig.de}

\author{Inti \surname{Sodemann Villadiego}\,\orcidlink{0000-0002-1824-5167}}
\email{sodemann@itp.uni-leipzig.de}
\affiliation{
Institut f\"ur Theoretische Physik,
Universit\"at Leipzig,
04103 Leipzig, Germany
}

\date{\today}

\maketitle


\beginsupplement
\tableofcontents

\section{Linear response for dielectric function}
\label{ap:linearresponse}


Consider a classical charged system specified by its charge density $\rho_c(\rr)$. 
In thermal equilibrium at inverse temperature $\be$, different charge configurations $\calC$ are distributed according to the Boltzmann weight $\prob(\calC)=\exp(-\be U_\calC)/\sum_\calC \exp(-\be U_\calC)$ where $U_\calC$ is the potential energy of the configuration. 
We will establish a sum-rule for the relative permittivity or dielectric constant $\eps$ of this system, treating $\eps^{-1}$ as a linear response of the infinite system to a perturbatively small external charge. 
A test charge $+q$ placed at the origin induces a change in the potential energy of a configuration $\calC$ by $\delta U=\int\dd{\rr} q\,v_C(r)\,\rho_c(\rr)$, where $v_C(r)$ is the pairwise Coulomb potential.  
This in turn leads to a change in the Boltzmann weights. To linear order the induced charge density is given by
    \begin{equation}\label{eq:deltacharge}
        \delta\!\ev{\rho_c(\rr)}
        =-\be\left[
        \ev{\rho_c(\rr)\delta U}_0
        -
        \ev{\rho_c(\rr)}_0\ev{\delta U}_0
        \right]
        =
        -\be q\int\dd{\rp} v_C(r')
        \left[
        \ev{\rho_c(\rr)\rho_c(\rp)}_0
        -
        \ev{\rho_c(\rr)}_0\ev{\rho_c(\rp)}_0
        \right],
    \end{equation}
where $\ev{\cdots}_0=\sum_{\calC}(\cdots)\prob(\calC)$ is the usual statistical average over the unperturbed configurations. 
The total potential at position $\rr$ is then the sum of the potentials set up by the test charge and the induced charge density,
    \begin{equation}\label{eq:Vtotal}
        V_\tx{tot}(\rr)
        =q\,v_C(\rr)
        +\int\dd{\rr''}v_C(\rr-\rr'')\,\delta\!\ev{\rho_c(\rr'')}.
    \end{equation}
We define the inverse dielectric function through $V_\tx{tot}(\rr)\equiv\int\dd{\rp}\eps^{-1}(\rr,\rp)\,v_C(\rp)q$. 
In the absence of translation and rotation invariance, $\eps^{-1}$ will generally be a function of two coordinates. 
Using Eqs.~\eqref{eq:deltacharge} and \eqref{eq:Vtotal} we immediately conclude that
    \begin{equation}\label{eq:eps_rrp}
        \eps^{-1}(\rr,\rp)=\delta(\rr-\rp)-\be\int\dd{\rr''}v_C(\rr-\rr'')\,C(\rr'',\rp),
    \end{equation}
where $C$ is the connected part of the two-point charge correlation function. 
In our present case of a Coulomb gas with fixed positive ions and particles of charge $q_i=-1$, the total charge density is $\rho_c(\rr)=\rho_{+}(\rr)-\rho(\rr)$, where $\rho_{+}(\rr)$ is the charge density of ions and $\rho(\rr)$ is the number density of itinerant particles. 
Since $\rho_{+}(\rr)$ is not dynamical, the charge correlation function in this case is $C(\rr,\rp)=S(\rr,\rp)$, where
$S(\rr,\rp)=\ev{\rho(\rr)\rho(\rp)}-\ev{\rho(\rr)}\ev{\rho(\rp)}$, is the connected piece of the electron density-density correlation function.

\section{Screening sum rules: relation to particle statistics}
\label{ap:sumrules}
At inverse temperature $\be=2m$, the (uniform) dielectric constant $\eps$ of a system of itinerant point charges (charge $-1$) can be defined as
    \begin{equation}\label{eq:eps_from_S}
        \eps^{-1}
        =1-4\pi m\lim_{q\to0}\frac{\avg{S}(q)}{q^2},
    \end{equation}
following the arguments in the main text. 
Here $\avg{S}(\qk)\coloneq\int\dd[2]{\rr}\e^{-\ii\qk\cdot\rr}\avg{S}(\rr)$ is the Fourier transform of the cell averaged correlation function, which is defined in terms of the local density operator $\rho(\rr)=\sum_{i=1}^N\delta(\rr-\rr_i)$ as
    \begin{equation}\label{eq:Sca}
        \avg{S}(\rr)%
        =\int\frac{\dd[2]{\xx}}{A}
        \left[
        \ev{\rho(\rr+\xx)\rho(\xx)}
        -
        \ev{\rho(\rr+\xx)}\ev{\rho(\xx)}
        \right].
    \end{equation}
From Eq.~\eqref{eq:eps_from_S}, it follows that $\avg{S}(q)$ starts at $\order{q^2}$. 
Assuming that $\avg{S}(\rr)$ has atleast $C_3$ symmetry, we can establish that in $2$D,
    \begin{equation}\label{eq:sumrule}
        \eps^{-1}=1+\pi m\int\dd[2]{\rr}r^2\avg{S}(\rr),
    \end{equation}
i.e. the density-density correlation function obeys a sum rule. This reduces to the Stillinger-Lovett second moment condition 
\cite{Martin.Gruber_JSP83_NewProofStillingerLovett,Martin_RMP88_SumRulesCharged} 
in case of perfect screening, $\eps^{-1}=0$. 
The existence of the limit in Eq.~\eqref{eq:eps_from_S} implies two other sum rules,
    \begin{equation}
        \int\dd[2]{\rr}\avg{S}(\rr)=0
        \qquad
        \tx{and}
        \qquad
        \int\dd[2]{\rr}r_\alpha\,\avg{S}(\rr)=0,\quad \alpha=x,y,
    \end{equation}
the first of which is a condition for number conservation.

We can connect the integral in Eq.~\eqref{eq:sumrule} to the statistics of particle positions in the state. 
Substituting Eq.~\eqref{eq:Sca} into the integral we get,
\begin{align}
    \int\dd[2]{\rr}r^2\avg{S}(\rr)
    &=\int\frac{\dd[2]{\xx}}{A}
        \left[
        \ev{\int\dd[2]{\rr}r^2\rho(\rr+\xx)\rho(\xx)}
        -
        \ev{\int\dd[2]{\rr}r^2\rho(\rr+\xx)}\ev{\rho(\xx)}
        \right]
    \no\\
    &=\sum_{i=1}^{N}
    \int\frac{\dd[2]{\xx}}{A}
        \left[
        \ev{(\rr_i-\xx)^2\rho(\xx)}
        -
        \ev{(\rr_i-\xx)^2}\ev{\rho(\xx)}
        \right]
    \no\\
    &=-2\sum_{i=1}^{N}
    \int\frac{\dd[2]{\xx}}{A}
        \left[
        \ev{\rr_i\cdot\xx\rho(\xx)}
        -
        \xx\cdot\ev{\rr_i}\ev{\rho(\xx)}
        \right]
    \no\\
    &=-\frac{2}{A}\sum_{i,j=1}^{N}
        \left[
        \ev{\rr_i\cdot\rr_j}
        -
        \ev{\rr_i}\cdot\ev{\rr_j}
        \right],
    \label{eq:positionvariance}
\end{align}
where we have repeatedly used the identities $\int\dd[2]{\xx}f(\xx)\rho(\xx)=\sum_if(\rr_i)$ and $\int\dd[2]{\xx}\rho(\xx)=N$.
Recasting Eq.~\eqref{eq:positionvariance} in terms of the ``center of mass'' coordinate, $\rr_\tx{CM}=(\sum_i\rr_i)/N$, we arrive at Eq.~(8) 
of the main text which connects $\eps^{-1}$ to the variance of $\rr_\tx{CM}$.

As a consequence of this relation, the usual Laughlin state in the LLL,
\begin{equation}\label{eq:laughlin}
        \Psi_0(\rr_1,\ldots,\rr_N)=\prod_{i<j}(\zee_i-\zee_j)^{m}\e^{-\sum_i\frac{|\rr_i|^2}{4\lb^2}},
    \end{equation}
has perfect long-distance screening, i.e. $\eps^{-1}=0$ at any filling. 
The exponent in Eq.~\eqref{eq:laughlin} can be exactly rewritten as
    \begin{equation}
        \sum_{i=1}^N|\rr_i|^2=\frac{1}{N}\sum_{i<j}|\rr_i-\rr_j|^2+N|\rr_\tx{CM}|^2,
    \end{equation}
which immediately implies that the variance of $\rr_\tx{CM}$ is $\lb^2/N=A_\tx{uc}/(2m\pi N)$ and hence $\eps^{-1}=0$.
\section{Density correlation function of a 2D OCP}

Consider a 2D model of classical charged point-like particles of charge $-1$ placed in a static background of ions of total charge $+1$ and radius $\sigma$. 
The ions are centered at lattice sites $\{\XX_i\}$ with lattice spacing $a$. 
We will consider a regime where $\sigma\ll a$ i.e. the ion charge is localized inside a small region compared to the size of the unit-cell. 
Each ion sets up a potential $-\chi(\rr-\ve{\rm X}_i)$, with
	\begin{equation}
		\chi(\rr)=
		\begin{dcases}
			\chi_{<}(\rr) & |\rr|<\sigma
			\\
			\ln(|\rr|/\sigma) & |\rr|>\sigma
		\end{dcases}
	\end{equation}
The form of the potential in the core region $|\rr|<\sigma$ depends on the details of the charge distribution in the core. For example if the charge density is uniform , $\chi_{<}(\rr)=(|\rr|^2-\sigma^2)/2\sigma^2$. 
We will derive the asymptotic scaling of the density correlation function $S(\rr,\rp)=\ev{\rho(\rr)\rho(\rp)}-\ev{\rho(\rr)}\ev{\rho(\rp)}$ in the limit of large separations $|\rr-\rp|\gg a,\sigma$. We will initially keep a general functional form for $\chi_<$ since it is not relevant for some parts of the derivation.

The point charges interact with each other via the 2D Coulomb potential $-\ln(|\rr|/L)$, where $L$ is a length-scale that sets the zero of the potential, and is chosen to be $L=\sigma$ for the rest of the discussion. The total potential energy of a configuration of charges $\{\rr_i\}$ is
	\begin{equation}
		\mathcal{U}=
		-\sum_{i<j}\ln\frac{|\rr_i-\rr_j|}{\sigma}
		+\sum_{i,j}\chi(\rr_i-\XX_j)
		-\sum_{i<j}\chi(\XX_i-\XX_j)
	\end{equation}
where the three terms are the charge-charge, charge-background, and background-background interactions. 
At low temperatures, the charges bind tightly to the positive ions, and their correlations become negligible. 
We perform a Mayer cluster expansion around this uncorrelated limit, i.e. the full Boltzmann weight of a configuration at inverse temperature $\be=2m$ is rewritten in terms of Mayer factors $f_{ij}\coloneq \e^{-2m\delta\mathcal{U}_{ij}}-1$, as
	\begin{equation}\label{eq:BoltzmanWeight}
		\e^{-2m\mathcal{U}}=\prod_k p(\rr_k-\XX_k)\prod_{i<j}(1+f_{ij})
		\quad\tx{where}\quad
		p(\rr)=\frac{1}{w}\e^{-2m\chi(\rr)},
        \quad
        w
		=\int_{|\xx|<\sigma}\dd[2]{\xx}\e^{-2m\chi_<(\xx)}+\frac{\pi\sigma^2}{m-1}.
	\end{equation}
The residual potential in the definition of the Mayer factor is
    \begin{equation}\label{eq:defDeltaU}
    \delta\mathcal{U}_{ij}(\rr_i,\rr_j)
    =-\ln\frac{|\rr_i-\rr_j|}{\sigma}
    -\ln\frac{|\XX_i-\XX_j|}{\sigma}
    +\chi(\rr_i-\XX_j)
    +\chi(\rr_j-\XX_i)
    \end{equation}
We will evaluate the density correlation function in a perturbation series in these Mayer factors, which themselves will be expanded to leading order in $\sigma/a$. 

\subsection{Leading order}
At leading order, the ion-electron pairs are uncorrelated. The density correlation function is $S^{0}(\rr,\rp)=\sum_ip(\rr_i-\XX_i)(\delta(\rr-\rp)-p(\rp-\XX_i))$, and it's cell averaged counterpart is
    \begin{equation}\label{eq:S0ca}
		\avg{S^0}(\dee)=\frac{1}{A_\tx{uc}}\,\delta(\dee)-\frac{1}{A_\tx{uc}}\int\dd[2]{\xx}\,p(\dee+\xx)\,p(\xx)
	\end{equation}
where $\dee=\rr-\rp$ is the separation and $A_\tx{uc}$ is the area of the unit cell. 
To find the asymptotic behavior of $\avg{S^0}(\dee)$ for $|\dee|\gg \sigma$, we split the integration domain in Eq.~\eqref{eq:S0ca} into three regions, 
	\begin{equation}
		\int\dd[2]{\xx}
        =\int_{x<\sigma}\dd[2]{\xx}
        +\int_{\substack{x>\sigma\\|\dee+\xx|<\sigma}}\dd[2]{\xx}
        +\int_{\substack{x>\sigma\\|\dee+\xx|>\sigma}}\dd[2]{\xx}
        .
	\end{equation}
To evaluate the asymptotics of the integral in these regions, we will make use of the following multipole expansion,
\begin{equation}
    \frac{\sigma^{2m}}{(a^2+b^2+2ab\cos\theta)^{m}}
    =\frac{\sigma^{2m}}{r_>^{2m}}
    \left[1
    -2m\cos\theta\frac{r_<}{r_>}
    +(m^2+m(m+1)\cos2\theta)\frac{r_<^2}{r_>^2}
    +\order{\frac{r_<^3}{r_>^3}}
    \right],
\end{equation}
where $r_>=\max(a,b)$ and $r_<=\min(a,b)$. 
In the first region, the integral goes as
    \begin{equation}
        I_1\approxeq \frac{1}{w}\frac{\sigma^{2m}}{d^{2m}}\int_{x<\sigma}\dd[2]{\xx}p(\xx)
        +
        \order{\frac{\sigma^{2m+2}}{d^{2m+2}}}
    \end{equation}
The integral in the second region is identical to that in the first after a variable change $\yy=-\dd-\xx$ and exploiting rotational symmetry. It has the same asymptotic behavior. 
The region in the third integral can be split up further,
    \begin{equation}
        \int_{\substack{x>\sigma\\|\dee+\xx|>\sigma}}\dd[2]{\xx}
        =
        \int_{\sigma<x<d-\sigma}\dd[2]{\xx}
        +
        \int_{x>d+\sigma}\dd[2]{\xx}
        +
        \int_{d-\sigma<x<d+\sigma}\dd[2]{\xx}
    \end{equation}
Only the first region has a contribution at the order of interest, the other two regions are subleading ($\order{d^{-(4m-2)}}$ and $\order{d^{-4m}}$ respectively). 
In the first region the integral behaves as
    \begin{equation}
        I_3=
        \frac{1}{w^2}\int_{\sigma}^{d-\sigma}\dd[2]{\xx}
        \frac{\sigma^{2m}}{x^{2m}}\frac{\sigma^{2m}}{|\dee+\xx|^{2m}}
        \approxeq
        \frac{1}{w}\frac{\sigma^{2m}}{d^{2m}}\int_\sigma^{d-\sigma}\frac{\dd[2]{\xx}\,\sigma^{2m}}{w\,x^{2m}}+\order{\frac{\sigma^{2m+2}}{d^{2m+2}}}
        =
        \frac{1}{w}\frac{\sigma^{2m}}{d^{2m}}
        \frac{\pi\sigma^2}{(m-1)w}
        +\order{\frac{\sigma^{2m+2}}{d^{2m+2}}}
    \end{equation}
Putting all the pieces together, we have
    \begin{equation}
        \avg{S^0}(\dee)\approx
        -\frac{1}{w\,A_\tx{uc}}\frac{\sigma^{2m}}{d^{2m}}\left[1+\int_{|\xx|<\sigma}\dd[2]{\xx}\,p(\xx)
        \right].
    \end{equation}
Note that at leading order, the connected correlator is $\order{\sigma^2/a^2}$ (in units of $1/\sigma^4$). 
\bibliography{LaughlinCrystal}